\documentclass{aa}  

\usepackage[varg]{txfonts}
\usepackage{graphicx}
\usepackage{gensymb}
\usepackage{appendix}
\usepackage{bigints}
\usepackage{placeins}
\usepackage{soul}
\usepackage{hyperref,natbib}
\hypersetup{
    colorlinks,
    citecolor=blue,
    linkcolor=blue,
    urlcolor=blue
}
\usepackage{amstext}
\usepackage{pifont}
\usepackage{multirow}
\usepackage{amssymb}
\usepackage[dvipsnames]{xcolor}

\usepackage{amsbsy}
\usepackage{caption}

\newcommand{\xmark}{\ding{55}}%

\begin{document} 

   \title{Sagittarius A* near-infrared flares polarization as a probe of space-time I:}
   \subtitle{Non-rotating exotic compact objects}

   \author{Nicolas Aimar\inst{1,2} \and Jo\~ao Lu\'is Rosa\inst{3.4}, Hanna Liis Tamm\inst{3}, Paulo Garcia\inst{1,2}}

   \institute{CENTRA, Faculdade de Engenharia, Universidade do Porto, s/n, R. Dr. Roberto Frias, 4200-465 Porto, Portugal\\
              \email{ndaimar@fe.up.pt}
              \and CENTRA, Departamento de Física, Instituto Superior Técnico-IST, Universidade de Lisboa-UL, Avenida Rovisco Pais 1, 1049-001 Lisboa, Portugal
              \and Institute of Physics, University of Tartu, W. Ostwaldi 1, 50411 Tartu, Estonia
              \and Departamento de Física Teórica, Universidad Complutense de Madrid, E-28040 Madrid, Spain
}

   \date{Received month xx, 2025; accepted month xx, 2025}
 
  \abstract
   {The center of our galaxy hosts Sagittarius~A*, a supermassive compact object of $\sim 4.3\times 10^6$ solar masses, usually associated with a black hole. Nevertheless, black holes possess a central singularity, considered unphysical, and an event horizon, which leads to loss of unitarity in a quantum description of the system. To address these theoretical inconsistencies, alternative models, collectively known as exotic compact objects, have been proposed.}
   {In this paper, we investigate the potential detectability of signatures associated with non-rotating exotic compact objects (ECOs) within the Sgr~A* polarized flares dataset, as observed through GRAVITY and the upcoming GRAVITY+.}
   {We examine a total of eight distinct metrics, originating from four different categories of static and spherically symmetric compact objects: Black Holes, Boson stars, Fluid spheres, and Gravastars. Our approach involves utilizing a toy model that orbits the compact object in the equatorial plane, at the Schwarzschild-Keplerian velocity. Using simulated astrometric and polarimetric data with present GRAVITY uncertainties as well as improved flux uncertainties expected for the GRAVITY+ upgrade, we fit the datasets across all metrics examined. We evaluated the detectability of the metric for each dataset based on the resulting $\chi^2_\mathrm{red}$ and BIC-based Bayes factors.}
   {Plunge-through images of ECOs affect polarization and astrometry in a distinguishable way from the spin of a Kerr black hole. With GRAVITY's present uncertainties, none of the metrics model are discernible. However, when the data are modeled within a compact Boson star background, the corresponding best-fit is sufficiently superior to the Kerr fit to rule out the latter. We examined the best expected enhanced flux uncertainties and discovered that a fourfold increase in flux sensitivity enables the detection of some of the exotic compact object models investigated. The others have a signal that is too close to each other to be distinguishable. However, with the GRAVITY+ flux uncertainties, when the data are produced using an ECO model, the best-fit ECO model is significantly preferred (with a BIC-based Bayes factor exceeding two) over the best-fit in the Kerr metric such that the latter can be ruled out. Nevertheless, enhancing the astrophysical complexity of the hot-spot model may diminishes these outcomes.}
   {With the improved sensitivity of GRAVITY+, we should be able to determine whether Sgr~A* is a Kerr black hole or some form of exotic compact object, although we will not be able to identify which specific ECO models provide the best description of Sgr~A*.}

   \keywords{Black hole physics - Relativistic processes - Polarization - Accretion, accretion disk - Radiative transfer}

\titlerunning{Sgr A* NIR flares polarization as probe of space-time I: Non rotating ECOs}
\authorrunning{Aimar, N; Lu\'is Rosa, J; Liis tamm, H; Garcia, P}

\maketitle
%
%-------------------------------------------------------------------

\section{Introduction}
The center of our galaxy hosts Sagittarius A* (Sgr~A*), a supermassive compact object of $(4.297 \pm 0.012) \times 10^6$ solar masses at a distance of only $8,277 \pm 9$ pc~\citep{gravity2022, gravity2023}. It is surrounded by star clusters, e.g. the so-called S-Star cluster, in which stars orbit around the compact object. The S-star proximity to Sgr~A* and their orbital parameters allowed to test General Relativity for supermassive compact objects, like the gravitational redshift for the star S2~\citep{gravity2018a} or the Schwarzschild precession~\citep{gravity2020a}. The orbit of these stars has been used to constrain the enclosed extended mass within the apocenter of S2 to be $\le$ 3000 $M_\odot$, that is, $\le 0.1\%$ of the mass of the supermassive compact object~\citep{gravity2022}. The S2 orbit has also been used to constrain the presence of scalar clouds~\citep{Foschi2023}, vector clouds~\citep{Foschi2024} and a fifth force~\citep{Gravity2025} around Sgr~A* without any significant evidence of their presence for scalar and vector clouds and with upper limits for the fifth force.
Indeed, the closest pericenter passage of the stars detected so far is still at a few thousand gravitational radii $r_g$, i.e. not in the strongest gravitational regime. The current observational state of the art cannot fully constrain the nature of the supermassive compact object at the center of the galaxy~\citep{deLaurentis_2023}.

Although the space-time around black holes can effectively describe the observations mentioned in the foregoing sections, these space-times present inherent difficulties from both mathematical and physical viewpoints. In essence, black hole space-times are characterized by singularities~\citep{Penrose1965, Penrose1969}, hinting at potential incompleteness within the theoretical framework. Moreover, the presence of an event horizon in black hole physics leads to the so-called black hole information paradox, in which the thermal nature of Hawking radiation implies a potential loss of information. This contradicts the principle of unitary evolution in quantum mechanics, where information must be preserved over time. This tension between general relativity and quantum theory has been widely discussed since Hawking's seminal work~\citep{Hawking1976}. To remedy such challenges, a number of alternative theoretical models, collectively termed exotic compact objects (ECOs), have been proposed (see Ref.~\citet{Cardoso2019} for a review). A subset of these ECO models can emulate similar observational predictions, thereby earning the designation of black hole mimickers.

The Event Horizon Telescope (EHT) has provided the first horizon-scale image of Sgr A*~\citep{EHT2022_I}. This landmark observation opens the door to testing the nature of compact objects in the strong gravity regime. While the image is broadly consistent with the predictions of general relativity for a Kerr black hole, it does not definitively rule out the existence of ECOs~\citep{EHT2022_VI, CarballoRubio2022, Vagnozzi2022, Shaikh2022, Ayzenberg2025}. Current resolution and modeling uncertainties allow for a range of ECO models—such as boson stars, gravastars, or wormholes—to remain compatible with the EHT data~\citep{GRMHDECO2024}.

Since 2001, outbursts of radiation called flares have been detected from Sgr~A* in X-rays~\citep{Baganoff2001, Nowak2012, Neilsen2013, Barriere2014, Ponti2015, Haggard2019}, near-infrared~\citep[NIR;][]{Genzel2003, Ghez2004, Hornstein2007, Hora2014} and radio~\citep{Yusef-Zadeh2006, Bower2015, Brinkerink2015}. In the past two decades, Sgr~A* flares have been the subject of intense observational campaigns and research, although no consensus has yet been reached on their physical origin. Indeed, multiple types of models exist for Sgr~A* flares including (among others) red noise~\citep{Do2009}, hot-spot~\citep{Genzel2003, Broderick2006, Hamaus2009}, ejected blob~\citep{Vincent2014}, star-disk interaction~\citep{Nayakshin2004}, disk instability~\citep{Tagger2006}, or magnetic reconnection~\citep{Aimar2023, Lin2023}. 

Thanks to interferometric measurements with the GRAVITY/VLTI instrument~\citep{Gravity}, the GRAVITY Collaboration reported the detection of orbital motion for Sgr~A* flares~\citep{gravity2018b, gravity2023} at $\sim 9$ gravitational radii ($r_g$) with a low inclination $i \sim 160$° compatible with the constraints from the~\citet{EHT2024} results of $i\approx 150$°. This detection brought important constraints on the modeling of Sgr~A* flares, favoring hot-spot and ejected blob models. However, the physical origin of Sgr~A* flares is still under debate.

Flares occur at only a few gravitational radii in the strong-field regime. They are thus an ideal object to study and constrain the space-time, and the nature of Sgr~A* (Schwarzschild, Kerr, or non-Kerr). Although the effects of spin or ECOs on the flare light curves are too small to be detected or degenerate with model parameters, the measure of the astrometry of flares with GRAVITY was thought to be sufficient to detect the non-Kerr metric signature. However, the low inclination and large uncertainties in the astrometric data do not allow for such a detection~\citep{Li2014a, Li2014b, Liu2015,Rosa:2023qcv}.

The detection of polarization of Sgr~A* flares in NIR~\citep{gravity2018b,gravity2023} and in radio~\citep{Wielgus2022} brought new observational constraints on the magnetic field configuration. In both wavelengths, the observed polarization properties (QU-loops and angular polarization velocity) are only compatible with a vertical magnetic field. \citet{Vincent2024} showed that the observed loop(s) in the QU-plane are mainly due to special relativity effects of the orbital motion of the emitting region. However, General Relativity, mainly light bending, also affects the observed QU-loop(s) creating an asymmetry relative to the horizontal axis~\citep{Vincent2024}. \textbf{In other words, linear polarization measured with Stokes parameters is sensitive to space-time curvature}.

Recently,~\citet{Rosa2025, tamm2025} studied the imprint on polarimetry of an orbiting hot-spot by non-rotating solitonic boson stars, gravastars and fluid spheres models. They show that the key difference in comparison with Schwarzschild relies on the presence and relative contribution from the additional images (second light ring or plunge-through images\footnote{Images formed when the geodesics pass through the interior of the compact object, which are absent in black hole spacetimes due to the presence of the event horizon.}) which mostly affect the polarization fraction, but also the Electric Vector Position Angle (EVPA).

The primary objective of this paper is to investigate the potential detectability of signatures associated with non-rotating ECOs within the Sgr~A* polarized flares dataset, as observed through the GRAVITY instrument. In Sect.~\ref{sec:metrics}, we discuss the various ECOs metric models that were taken into consideration for this study, while Sect.~\ref{sec:flare_model} elaborates on the flare model itself. The methodological framework is delineated in Sect.~\ref{sec:methodology}, followed by the presentation of our findings in Sect.~\ref{sec:results}. We discuss our results in Sect~\ref{sec:discussion} and made a summary accompanied by the conclusion in Sect.~\ref{sec:conclusion}.

\section{Horizonless Exotic Compact Objects metrics}\label{sec:metrics}
We analyze three types of static and spherically symmetric exotic compact objects whose optical properties have been previously analyzed with GYOTO in the context of infrared flares and radio (EHT) imaging. These are the solitonic Boson star, the relativistic perfect-fluid sphere, and the gravitational vacuum star (Gravastar). Table~\ref{tab:metric_params} summarizes the parameters of the models. All the selected models have a shadow that mimics the one of a black hole and are compatible with EHT observations~\citep[see][ for a review]{Cardoso2019}. 
%, and the black-hole surrounded by a perfect fluid dark matter halo~\citep{Macedo:2024qky}. 
%More information about these models can be found in the references above. In the following, for self-consistency, we summarize the fundamentals of each of these models.

\subsection{Solitonic Boson star}
Scalar Boson stars consist of localized solutions of self-gravitating scalar fields and have been the subject of intense theoretical effort~\citep{Kaup1968, Ruffini1969, Colpi1986, Friedberg1987, Jetzer1992, Schunck2003, Liebling2012, Macedo2013, Grandclement2017, Cunha2023}. They are found as solutions of the Einstein-Klein-Gordon theory described by the action
\begin{equation}
    S=\int\sqrt{-g}\left[\frac{R}{16\pi}-\frac{1}{2}\partial_\mu\phi^*\partial^\mu\phi-\frac{1}{2}V\left(|\phi|^2\right)\right]d^4x,
\end{equation}
where $R$ is the Ricci scalar, $g$ is the determinant of the metric $g_{\mu\nu}$, $\phi$ is the complex scalar field, with ${\ ^*}$ denoting complex conjugation and $|\phi|^2=\phi^*\phi$, and $V$ is the scalar potential. Different Boson star models are obtained depending on the form of the potential $V$. In particular, solitonic Boson stars are described by a potential of the form $V=\mu^2|\phi|^2\left(1-|\phi|^2/\alpha^2\right)^2$, where $\mu$ is a constant that plays the role of the mass of $\phi$ and $\alpha$ is a constant free parameter of the model~\citep[see][ for more details]{Lee1987, Rosa2022}. The main interest behind solitonic Boson stars lies in the fact that these can be compact enough to develop bound-photon orbits while maintaining their stability against radial perturbations. 

\begin{figure}
    \centering
    \includegraphics[width=\linewidth]{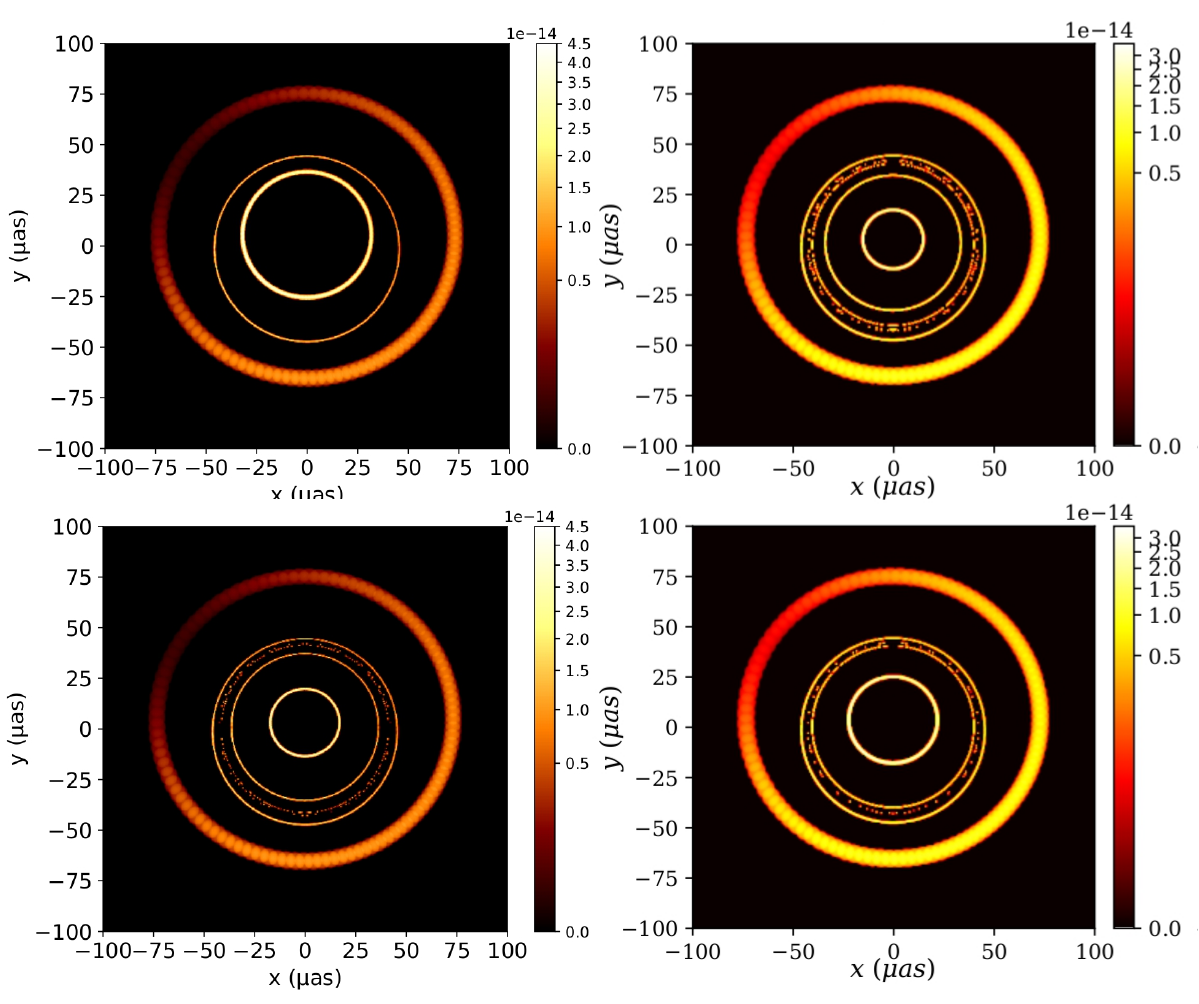}
    \caption{Time integrated image of a hot-spot orbiting the Boson star 2 (\textbf{top-left}), Boson star 3 (\textbf{bottom-left}), Fluid sphere 2 (\textbf{top-right}) and Fluid sphere 3 (\textbf{bottom-right}) models with an inclination close to face-on ($i=20^\circ$). Extracted from~\citet{Rosa2025} and~\citet{tamm2025}.}
    \label{fig:BosonStar_Fluid_Sphere}
\end{figure}

Due to the complexity of the Einstein-Klein-Gordon system of field equations, no analytical solutions describing Boson stars have been obtained. We consider two numerical solutions of solitonic Boson stars with different compactness, a model close to the ultra-compact regime (Boson star 2), and an ultra-compact model (Boson star 3) described in detail in~\citet{Rosa:2023qcv}. We exclude the dilute model (Boson star 1) as the observed Q-U loops for this model are large~\citep{Rosa2025} which is incompatible with the GRAVITY data~\citep{gravity2023}. Time integrated images of an orbiting hot-spot at low inclination around these two configurations are shown in the left panels of Fig.~\ref{fig:BosonStar_Fluid_Sphere}. The Boson star 2 model shows a large bright and thick inner ring corresponding to the plunge-through image, while the Boson star 3 model shows a pair of light-rings and a smaller, but still bright, inner ring, again corresponding to the plunge-through image.

\subsection{Relativistic Fluid sphere}
Relativistic Fluid spheres are solutions to Einstein's field equations in the presence of an isotropic perfect fluid~\citep{Tolman1939, Oppenheimer1939, Buchdahl1959, Iyer1985}. The interior of these solutions is described in the usual spherical coordinates $(t, r, \theta, \phi)$ by the metric
\begin{eqnarray}
        ds^2&=&-\frac{1}{4}\left(3\sqrt{1-\frac{2M}{R}}-\sqrt{1-\frac{2r^2M}{R^3}}\right)^2dt^2+\\
        &+& \left(1-\frac{2r^2M}{R^3}\right)^{-1}dr^2+r^2\left(d\theta^2+\sin^2\theta d\phi^2\right),\nonumber
\end{eqnarray}
where $M$ and $R$ are constants that represent the mass and the surface radius of the star, respectively. The exterior of these solutions is described by the Schwarzschild solution. We restrict our analysis to smooth solutions, that is, the surface of the star $R$ coincides with the radius at which the matching between the interior and exterior solutions is performed, and with a constant volumetric energy density. Under these restrictions, these solutions develop a pair of bound-photon orbits whenever $R\leq3M$, with the limiting case $R=3M$ corresponding to a single degenerate bound-photon orbit, whereas a singularity is present whenever $R\leq2.25M$. More details about this model can be found in Ref.~\citet{Rosa2020, Tamm:2023wvn}. Three models were studied in~\citet{tamm2025} with $R=2.25 M$ (Fluid sphere 1), $R=2.5 M$ (Fluid sphere 2), and $R=3 M$ (Fluid sphere 3), however, at low inclination (which is likely the case for Sgr~A*), the Fluid sphere 1 model presents the same observational features as Schwarzschild. Thus, in this study, we consider only the Fluid sphere 2 and Fluid sphere 3 models. Similarly to the Boson star case, the right panels of Fig.~\ref{fig:BosonStar_Fluid_Sphere} show the time integrated images of a hot-spot orbiting the Fluid sphere 2 (top row) and Fluid sphere 3 (bottom row) models at low inclination. Both of them have a pair of light rings and a small inner ring corresponding to the plunge-through image. The Fluid sphere 2's inner ring is smaller than the Fluid sphere 3 model. Thus, its contribution to the polarization is smaller. 

\subsection{Gravastar}
\begin{figure*}
    \centering
    \includegraphics[width=\linewidth]{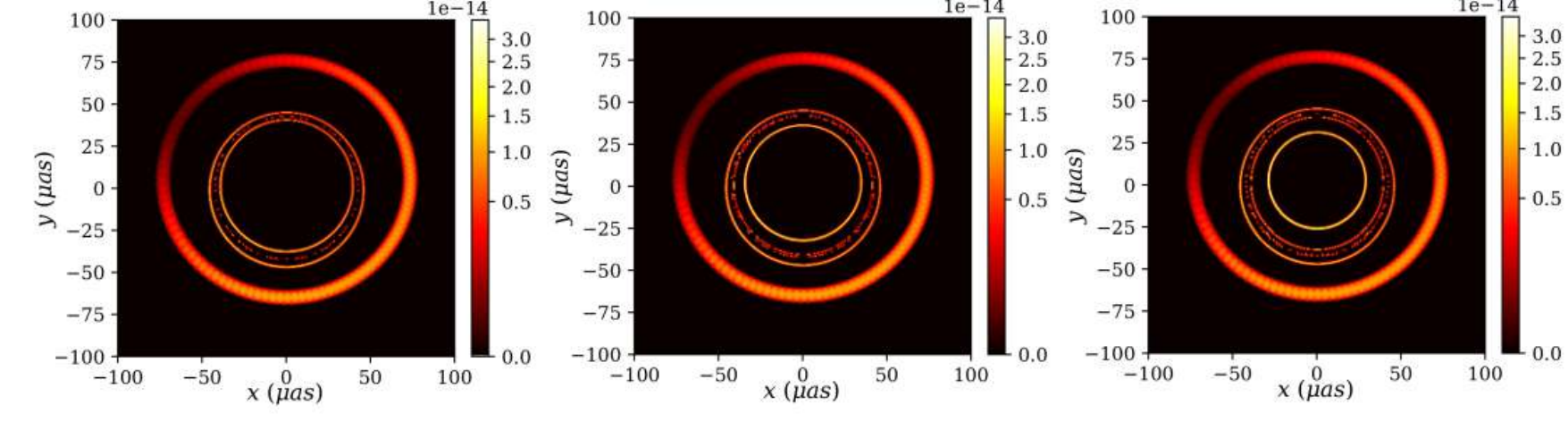}
    \caption{Time integrated image of a hot-spot orbiting the Gravastar models with $R=3M$ (Gravastar~1) in the \textbf{left panel}, $R=2.5M$ (Gravastar~2) in the \textbf{middle panel} and $r=2.01M$ (Gravastar~3) in the \textbf{right panel}. Extracted from~\citet{tamm2025}.}
    \label{fig:GravaStar}
\end{figure*}

Similarly to relativistic Fluid spheres, gravitational vacuum stars, or Gravastars, are solutions of the Einstein's field equations in the presence of an isotropic perfect fluid. In the case of Gravastars, this fluid is exotic, satisfying an equation of state of the form $p=-\rho$, where $p$ is the isotropic pressure and $\rho$ is the energy density~\citep{Mazur2004, Visser2004, Mottola2006, Pani2009, Mazur2015, Danielsson2018, Posada2019, Mazur2023}. The interior of Gravastars is described by the metric
\begin{equation}
    ds^2=-\alpha\left(1-\frac{2m(r)}{r}\right)dt^2+\left(1-\frac{2m(r)}{r}\right)^{-1}dr^2+r^2\left(d\theta^2+\sin^2\theta d\phi^2\right),
\end{equation}
where $\alpha$ is a parameter controlling the volumetric mass distribution of the star, and $m(r)=\frac{4}{3}\pi \rho r^3$ is the mass function. The exterior of the Gravastar is described by the Schwarzschild space-time. The parameter $\alpha$ is bounded between $1$ and a minimum value $\alpha_{\text{min}}$ that depends on the model, with $\alpha=1$ and $\alpha=\alpha_{\text{min}}$ corresponding to solutions with all mass distributed in volume or on the surface, respectively. Regardless of the mass distribution, these solutions develop a pair of bound-photon orbits whenever the radius $R$ of the surface of the Gravastar, where the matching between the interior and exterior space-times is performed, satisfies the condition $R\leq 3M$, with the limiting case $R=3M$ corresponding to a single degenerate bound-photon orbit. More details about this model can be found in Ref.~\citet{Rosa:2024bqv}. For $\alpha \neq 1$, the $g_{rr}$ coefficient of the metric is discontinuous, making the parallel transport of the polarization basis inside the ray tracing code \texttt{GYOTO} (more details in Section~\ref{sec:gyoto}) impossible. We thus fixed $\alpha=1$ and selected three configurations: a model with $R=3M$~(Gravastar~1), $R=2.5M$~(Gravastar~2) and $R=2.01M$~(Gravastar~3). Again, we present the time integrated images of a hot-spot orbiting these three configurations at low inclination in Fig.~\ref{fig:GravaStar}. All configurations present a set of light-rings and at least one plunge through image. The size of the inner plunge-through image decreases with an increase in compacticity, with the most compact configuration (Gravastar~3) presenting an additional light ring.
%The least compact configuration, referred to as the Gravastar~1 model, exhibits only a pair of light rings without any plunge-through images, whereas the other two models display both a pair of light rings and plunge-through images, with the latter ring that varies in intensity and size.

\begin{table}[]
    \centering
    \begin{tabular}{lccc}
        \hline
        \hline
        Metric model & R [$r_g$] & Compactness & \# of light ring(s)\\[0,1cm]
        \hline
        Schwarzschild & 2 & 0.5 & 1 \\[0,1cm]
        Boson star 2 & 3.95 & $\sim 0.253$ & 1 \\[0,1cm]
        Boson star 3 & 3.12 & $\sim 0.321$ & 2 \\[0,1cm]
        Fluid sphere 2 & 2.5 & 0.4 & 2 \\[0,1cm]
        Fluid sphere 3 & 3 & $\sim 0.333$ & 2 (degen.) \\[0,1cm]
        Gravastar~1 & 3 & $\sim 0.333$ & 2 (degen.) \\[0,1cm]
        Gravastar~2 & 2.5 & 0.4 & 2 \\[0,1cm]
        Gravastar~3 & 2.01 & $\sim 0.498$ & 2  \\[0,1cm]
        \hline
    \end{tabular}
    \caption{Summary of the metric models properties.}
    \label{tab:metric_params}
\end{table}

\section{Hot-spot model} \label{sec:flare_model}
We consider an analytical orbiting hot-spot model for the flares of Sgr~A* in NIR as it is the model that agrees the most with the current data. However, we did not choose any specific physical phenomenon at the origin of the flare to avoid model dependence. The hot-spot is assumed to be a uniform sphere of plasma with a radius of $1\, r_g$ that emits synchrotron radiation.

\begin{figure}
    \centering
    \includegraphics[width=\linewidth]{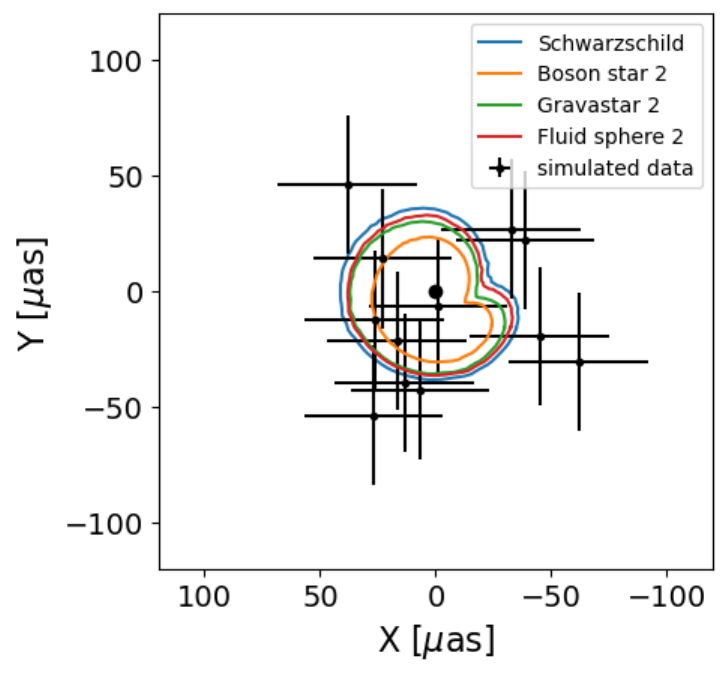}
    \caption{Astrometry of an orbiting hot-spot in different metric models and simulated astrometric data with current GRAVITY uncertainties.}
    \label{fig:astrometry}
\end{figure}

As previously demonstrated~\citep{Li2014a, Liu2015, Rosa:2023qcv} and depicted in Fig.~\ref{fig:astrometry}, the current uncertainties in astrometry, alongside the fact that the flares from Sgr~A* are observed nearly face-on, prevent the differentiation between various ECO and Schwarzschild models, emphasizing the necessity for polarization measurements.

\subsection{Polarized ray-tracing} \label{sec:gyoto}
We used the public polarized ray-tracing code \texttt{GYOTO}\footnote{\url{https://github.com/gyoto/Gyoto}}~\citep{gyoto2011, gyoto2024} to compute the images of the hot-spot model for the four Stokes parameters (I, Q, U and V) which characterize the total received intensity (Stokes I) and the polarization of the received light (Q and U for linear polarization and V for circular polarization). For this study, we ignored the circular polarization V as it is not measurable by GRAVITY.

For each observing time, we compute the field integrated polarized fluxes $F_I(t)$, $F_Q(t)$, and $F_U(t)$ in Jansky, and the centroid position of the total flux ($X(t), Y(t)$) from the computed images. While polarization strongly depends on the orbital parameters and the inclination (see below), the incorporation of the astrometry allows for breaking some degeneracies.
For the rest of the paper, the field of view is set to 2.6 times the orbital radius in M units to optimize the computation time, and the default resolution is 300x300 pixels. This resolution is sufficient to get the low-order plunge-through images from the ECO models but not the high-order ones, which require an extreme resolution but have a minor impact on the observed fluxes.
The observed wavelength is set to the one of GRAVITY, i.e., $2.2\, \mu$m.

\subsection{Orbital motion}
For this theoretical study, we choose the simplest type of orbital motion, i.e., a circular orbit at radius $r$ in the equatorial plane. The velocity of Sgr~A*'s IR flares as well as the spin of Sgr~A* are not well constrained because of the large astrometric uncertainties. Some recent studies are suggesting a possible super-Keplerian motion~\citep{gravity2018b, Aimar2023, Antonopoulou2024, Yfantis2024a, Yfantis2024b, Xie2025}. Moreover, as shown in Fig.~\ref{fig:astrometry}, the astrometric tracks in the ECOs metric are smaller for the same orbital radius due to the contribution of the inner plunge-through images. Thus, when we want to fit simulated data of a Keplerian orbit in an ECO metric (see more details in Sect.~\ref{sec:methodology}) using the Kerr metric, the apparent orbital radius will be smaller, and the velocity will be sub-Keplerian. On the contrary, when fitting simulated data of a keplerian orbit in Kerr using an ECO metric, the apparent orbit will be larger with a super-Keplerian motion. Thus, we allow the orbital velocity to be a free parameter in the form of a coefficient $K_\mathrm{coef}$ which is the fraction of the Schwarzschild Keplerian velocity. The orbital motion is thus defined in Boyer-Lindquist coordinates as
\begin{subequations}
    \begin{align}
        \theta &= \frac{\pi}{2}, \\
        \frac{dr}{dt} &= 0, \\
        \frac{d\theta}{dt} &= 0, \\
        \frac{d\varphi}{dt} &= K_\mathrm{coef} \cdot \left(r^{3/2} + a \right)^{-1}, \label{eq:orbital_velo}
    \end{align}
\end{subequations}
with $a$ the spin of the compact object (only for the Kerr metric in this study).

\subsection{Electron Energy Distribution function}
The IR flux from Sgr~A* flares is thought to be generated by synchrotron radiation from a non-thermal population of electrons~\citep{gravity2021}. We thus consider a $\kappa$-distribution of electrons~\citep{Pandya2021} for our hot-spot. This distribution is characterized by a thermal core with a temperature $T_e$ and a power-law tail with $\kappa$-index\footnote{related to the power law index $p$ as $\kappa = p+1$} and a number density $n_e$. These three parameters have a strong influence on the fluxes, while the $\kappa$-index governs the NIR spectral index. To simplify and limit the number of parameters, we set the number density to $10^7$ cm$^{-3}$, the temperature to $5\times 10^{10}$K and the $\kappa$-index to $3.5$. The overall flux of the generated synthetic flares with these settings matches the usual observed flux from Sgr~A* flares ($\approx10$ mJy). Obviously, not all flares have the same maximum flux. To mitigate this, we restricted the study to the normalized polarized fluxes $Q/I$ and $U/I$. 

\subsection{Magnetic field configuration}
The average of polarization measurements of multiple flares made by GRAVITY shows a single loop of polarization over one orbital period of $P=60\, $min which constrains both the magnetic field configuration to be vertical and the inclination to $i=157\degree \pm 5\degree$~\citep[see Fig. 4 of]{gravity2023}. This magnetic field configuration has also been investigated for ECO-metrics~\citep{Rosa2025, tamm2025}. We thus restrict our study to this magnetic field configuration. The magnetic field strength is defined through the magnetization parameter $\sigma = B^2 / 4 \pi m_p c^2 n_e$ fixed to 0.01. The magnetic field strength is thus around 43 G which is the expected order of magnitude for Sgr~A* flares~\citep{vonFellenberg2025}. Note that in this model, this parameter is only a scaling factor of the total flux.

The observed linear polarization fraction of Sgr~A* flares is $\sim 10-40\%$~\citep{gravity2020b}, much lower than the linear polarization fraction expected from a purely ordered magnetic field (without any stochastic component). A probable explanation is that the magnetic field is only partially ordered, i.e. with a stochastic component, as suggested by the EHT observation of Sgr~A*~\citep{EHT2024}. This effectively reduces the observed polarization fraction~\citep{Rybicki&Lightman1979}. The contribution of plunge-through images can also result in a decrease of the linear polarization fraction, but only for a specific part of the orbit and cannot be responsible for the global difference of polarization fraction. To take into account the low observed polarization fraction compared to the models, we apply a constant scaling factor $lp$ to the modeled polarized quantities (Stokes Q and U) from \texttt{GYOTO}, whose value is in [0,1]\footnote{This parameter should not be interpreted as the degree of order of the magnetic field as the relation is indirect and non-trivial.}.

\subsection{Parameters summary}\label{sec:params_summary}
Our model has a total of six free parameters. The hot-spot has five free parameters, summarized in Tab.~\ref{tab:hotspot_params}: inclination $i$, orbital radius $r$, initial azimuthal angle $\varphi_0$, orbital velocity through the fraction of the Keplerian velocity coefficient $K_\mathrm{coef}$ and the polarization factor $lp$. The last parameter is the background metric (metric model). We fixed the Position Angle of the Line of Nodes to 177.3$^\circ$ following~\citet{gravity2023} as this parameter add an unnecessary degree of freedom for this theoretical study.

The effect of the spin of a Kerr black hole on the polarization measurements $Q/I$ and $U/I$ could be similar to the impact of the plunge-through images from the ECOs metrics. Thus, to test if an ECO signature can be fitted and thus mimic by the spin of a Kerr black hole, we add the spin as a free parameter in the Kerr metric (see \ref{sec:test_spin}).

\begin{table}[h]
    \centering
    \begin{tabular}{lcc}
        \hline
        \hline
        Parameter & Symbol & Default value\\[0,1cm]
        \hline
        Orbital radius [$r_g$] & $r$ & $8.2$\\[0,1cm]
        Azimuthal angle at $t^{BL}=0$ [°] & $\varphi_0$ & $15$\\[0,1cm]
        Inclination [°] & $i$ & $157$\\[0,1cm]
        Orbital velocity scaling factor & $K_\mathrm{coef}$ & 1.0 \\[0,1cm]
        Linear polarization factor & $lp$ & $0.4$\\[0,1cm]
        \hline
    \end{tabular}
    \caption{Summary of the parameters of the hot-spot model.}
    \label{tab:hotspot_params}
\end{table}

\section{Methodology}\label{sec:methodology}
To determine the detectability of the metrics considered in this paper, we performed fits of simulated data with all aforementioned metrics and compared the results of these fits. Using statistical criteria, we conclude on the detectability of the metrics. The following section describes in detail our methodology.

\begin{figure*}
    \centering
    \includegraphics[width=\linewidth]{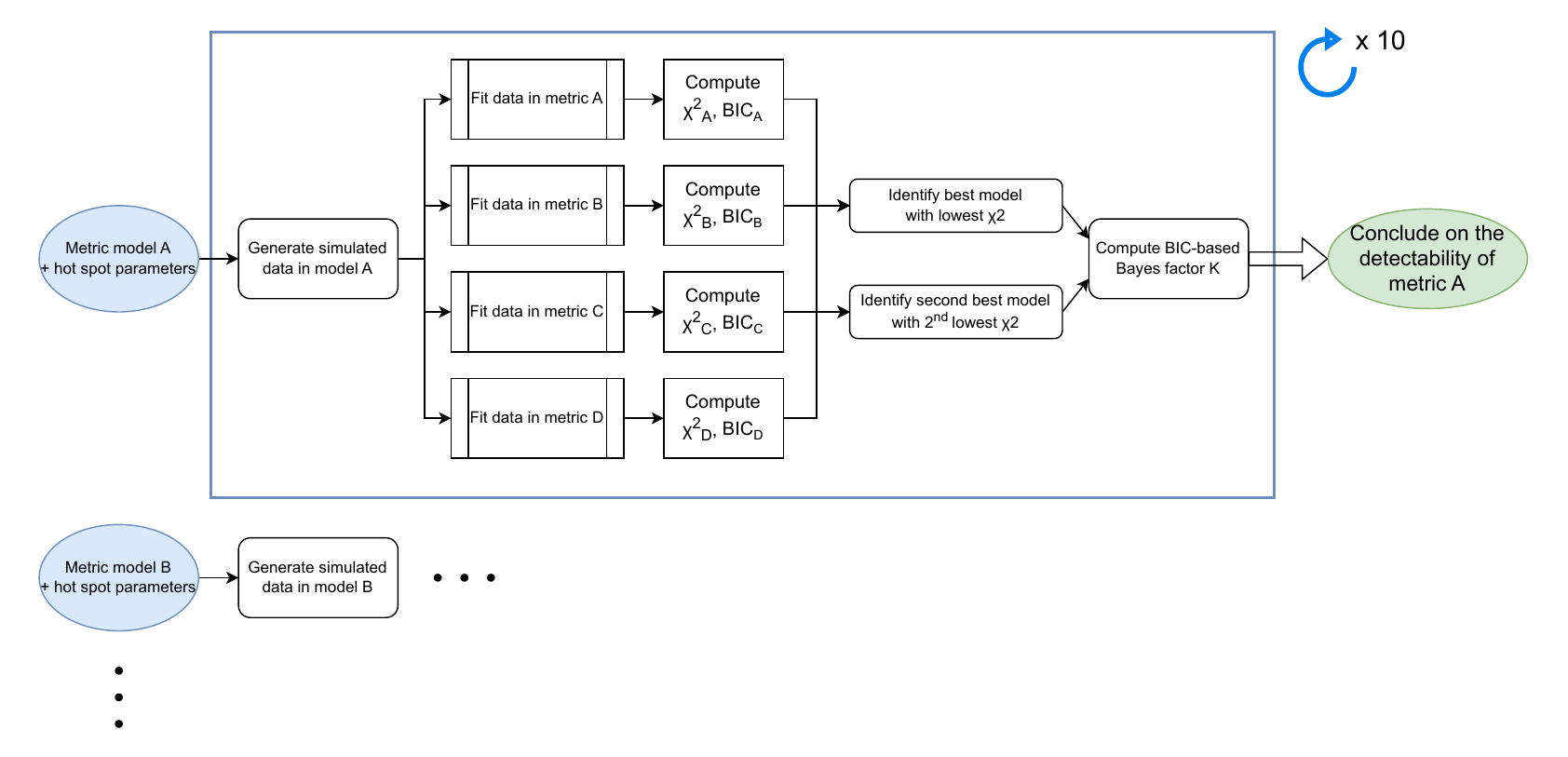}
    \caption{Sketch illustrating our methodology to assess the detectability of ECO's models.}
    \label{fig:sketch_method}
\end{figure*}

\subsection{Generation of simulated data}
We generated simulated data following the set of parameters presented in Tab.~\ref{tab:hotspot_params} for the hot-spot within a given background metric.

In order to accommodate the uncertainties inherent to observational data, we introduce a random variable to the simulated data points, which is drawn from a Gaussian distribution. For the simulated data on astrometric measurements, the Gaussian distribution is characterized by a standard deviation denoted $\sigma_\mathrm{Astro}=30\, \mu$as. For the simulated polarized flux measurements (I, Q and U), the standard deviation introduced is flux-dependant as 
\begin{equation} \label{eq:sigma_flux}
    \sigma_\mathrm{Flux}^\mathrm{GRAVITY} = 0.15 \cdot I^{0.6},    
\end{equation}
with $\sigma_\mathrm{Flux}^\mathrm{GRAVITY}$ and $I$ in mJy. The flux dependency of Eq.~\ref{eq:sigma_flux} has been reported by~\citet{gravity2020c} and here calibrated with the uncertainties from~\citep{gravity2023}\footnote{The uncertainties in \citep{gravity2023} are two times smaller than the one predicted by~\citet{gravity2020c}. We thus applied a factor 0.5 to their fitted equation.}. The uncertainties associated with the ratios $Q/I$ and $U/I$ are derived by propagating the measurement errors in $I$, $Q$, and $U$. In Sect.~\ref{sec:results} we investigate the detectability of the ECOs with various flux uncertainties in the form of fraction of the Eq.~\ref{eq:sigma_flux}, from 1 being the GRAVITY's actual uncertainties until $\sigma_\mathrm{Flux}^\mathrm{GRAVITY}/7$ which correspond to the best anticipated improved precisions for GRAVITY+ upgrade~\citep{Bourdarot2024}.

The observational duration and data point intervals are by default calibrated to resemble those of GRAVITY, namely, one measurement every 5 minutes resulting in a total of 12 observations, amounting to 60 minutes overall equivalent to one full orbital period. In Sect.~\ref{sec:results}, we also investigated the impact of a better time sampling on the detectability of the metrics.

\subsection{Fitting}
\subsubsection{Fitting procedure}
Contrary to the five hot-spot parameters, the metric parameter is not continuous (a range of values) but is instead categorical, i.e., a choice of a configuration. The treatment of this parameter is thus different from the others. As part of our fit strategy, we opted to individually fit the five (six for Kerr) remaining parameters of our model, namely, the inclination, the orbital radius, the initial azimuthal angle, the orbital velocity scaling factor and the polarization factor (plus the spin in Kerr), within the context of all possible ECO + Kerr background metrics. 

We perform a classical $\chi^2$ minimization using the \texttt{scipy.least\_square} algorithm with the \textit{Trust Region Reflective} method. This combination is selected because it is well suited to large, bounded optimization problems, which corresponds to our situation involving five or even six parameters with clearly defined limits. Another important aspect of the fitting strategy is the computational cost. Each model evaluation requires several minutes, so a single fit typically ranges from about 2 to 12 hours, depending on the chosen metric. This type of algorithm is highly dependent on the initial guess. To reduce the risk of converging to a local minimum, we adopt as our \textit{first-guess} the point with the minimum reduced chi-square, $\chi^2_\mathrm{red}$, obtained from a preliminary grid search (see Appendix~\ref{ap:grid} for details on the grid).
We evaluate the $\chi^2_\mathrm{red}$ as
\begin{equation}
    \chi^2_\mathrm{red} = \frac{1}{\mathrm{\mathrm{\nu}}}\sum_{i} \left( \frac{x_i-\mu_i}{\sigma_i} \right)^2,   
\end{equation}
where $\mathrm{\nu} = n-k$ is the number of degrees of freedom, $n$ being the number of data points and $k$ the number of free parameters, $x_i$ the data point $i$, $\mu_i$ the model and $\sigma_i$ the uncertainty of data point $i$. We also compute the \textit{Bayesian Information Criterion} (BIC) as~\citep{Kass1995}
\begin{equation}
    \mathrm{BIC} = \chi^2_\mathrm{red} \cdot \nu + k \ln \left(n \right)
\end{equation}
to obtain the statistical criteria of the different fits to be compared to the others.

\subsubsection{Evaluate the detectability of a metric}\label{sec:detectability}
To assess detectability, the primary requirement is that the best-fitting metric—that is, the one yielding the lowest reduced chi-square, $\chi^2_\mathrm{red}$—matches the background metric used to generate the simulated data. Yet, the impact of the plunge-through images (induced by the background metric) may not be large enough to result in a significantly better fit in terms of $\chi^2_\mathrm{red}$ for one metric relative to the others, or it may mimic the effect of the spin of a Kerr black hole (or vice versa) and thus lead to comparable $\chi^2_\mathrm{red}$ values. For this reason, we additionally employ the BIC-based Bayes factor $K$ between the best-fitting metric and the second-best-fitting metric, i.e. the one with the second-lowest $\chi^2_\mathrm{red}$. The BIC-based Bayes factor is evaluated following~\citet{Wagenmakers2007} as
\begin{equation}
    \log_{10} K = \log_{10} \left[ \exp \left(\frac{\Delta \mathrm{BIC}}{2} \right) \right],
\end{equation}
with $\Delta \mathrm{BIC}$ the difference of $\mathrm{BIC}$ between the second-best-fitting and best-fitting metrics.

For each background metric and uncertainty levels, we generated 10 simulated datasets. Even limited, this amount of simulated dataset allows deriving statistics (average and standard deviation) on the lowest $\chi^2_\mathrm{red}$, the associated $\mathrm{BIC}$ and more importantly $\log_{10} K$ with a limited computational cost. We thus define four detectability outcomes according to the value of $\log_{10} K$ and following the \citet{Kass1995} scale:

\begin{itemize}
    \item \textbf{risk of mismatch} : The best-fitting metric does not always correspond to the background metric in the simulated data, meaning that there is a risk of mismatch of metrics (symbolized by \textcolor{red}{\xmark}\textcolor{red}{\xmark}).
    \item \textbf{not detectable} : $\log_{10} K < 1$. The best-fitting metric corresponds to the background metric in the simulated data, but the second-best-fitting metric is equivalent to the best-fitting metric (symbolized by \textcolor{red}{\xmark}).
    \item \textbf{partially detectable} : $1 \leq \log_{10} K \leq 2$ . In this case, the best-fitting metric is significantly better, but not enough to make a strong statement (symbolized by \textcolor{orange}{\boldmath$\sim$\unboldmath}).
    \item \textbf{detectable} : $\log_{10} K > 2$. The best-fitting metric is considerably better to the second-best-fitting metric and is thus detectable (symbolized by \textcolor{ForestGreen}{\pmb{\checkmark}}).
\end{itemize}

A sketch summarizing the whole procedure is shown in Fig.~\ref{fig:sketch_method}.

Although some ECO metrics are difficult to distinguish from one another, it is also interesting to examine whether, even not been detectable, a metric can be discriminated from the Kerr solution. To this end, we calculate the BIC-based Bayes factor between the best-fitting metric and Kerr, denoted as $\log_{10} K|_\mathrm{Kerr}$. We adopt the same notation used for the detectability analysis when assessing the exclusion of the Kerr metric.

\section{Results} \label{sec:results}
\subsection{Current GRAVITY uncertainties}\label{sec:current_uncertainties}

\begin{figure}[h]
    \centering
    \includegraphics[width=\linewidth]{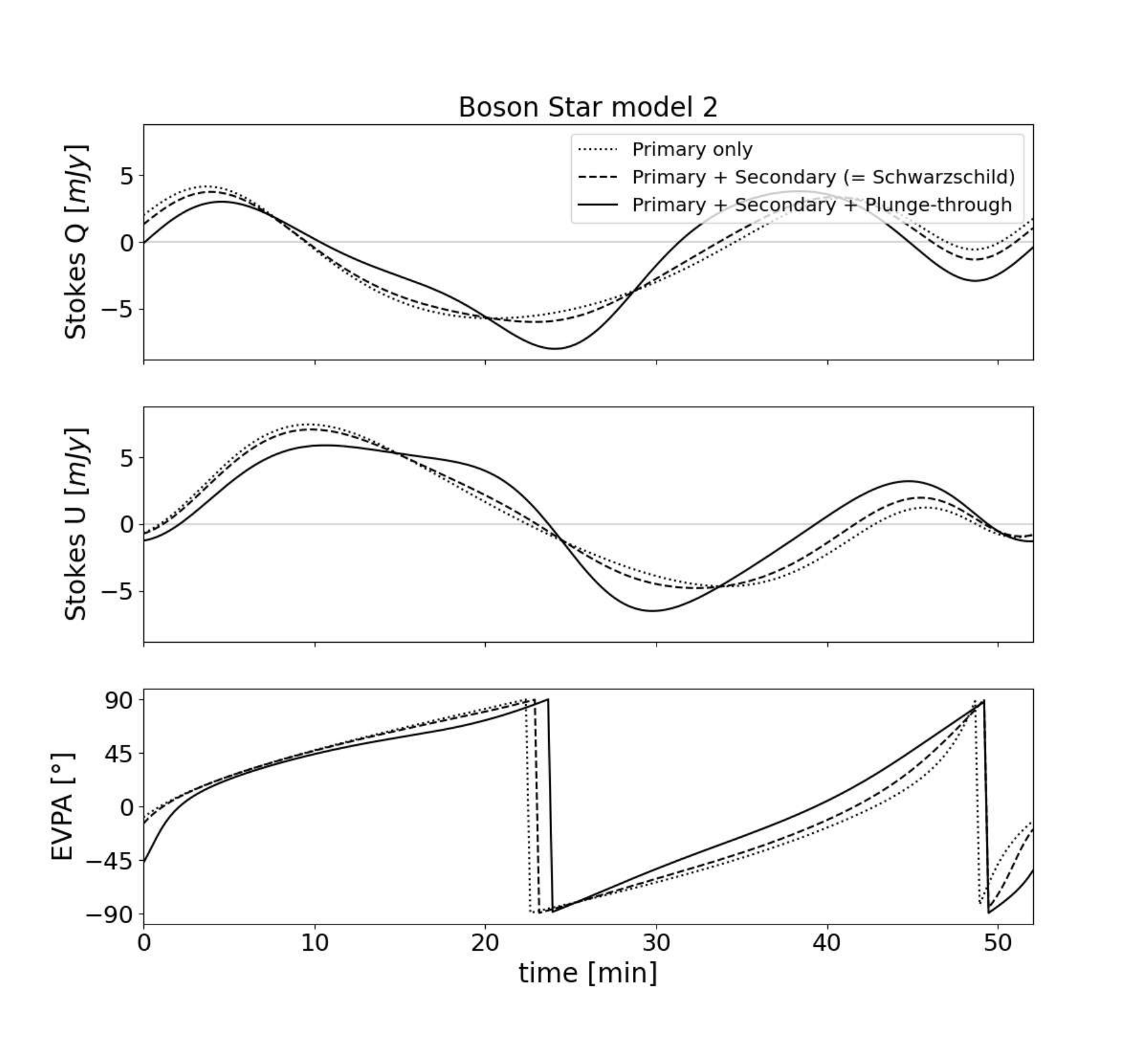}
    \caption{Contribution of the various image orders and nature, i.e. primary only in dotted lines, primary + secondary (equivalent to Schwarzschild) in dashed lines, and all images including the plunge-through images in solid lines.}
    \label{fig:BS2_Contrib}
\end{figure}

\begin{table*}[ht]
    \centering
    \begin{tabular}{l|c c c c c}
        \hline
        \hline
        Data's metric & $< \chi^2_\mathrm{red}>$ & $<BIC>$ & $<\log_{10} K>$ & detectability & $<\log_{10} K|_\mathrm{Kerr}>$ \\[0,1cm]
        \hline
        Schwarzschild & $0.94 \, \pm \, 0.14$ & $65 \, \pm \, 7$ & $-0.04 \, \pm \, 0.52$ & \textcolor{red}{\xmark}\textcolor{red}{\xmark} & $0.59 \, \pm \, 0.68$ \textcolor{red}{\xmark} \\[0,1cm]
        Boson star 2 & $0.92 \, \pm \, 0.22$ & $63\, \pm \, 11$ & $0.73 \, \pm \, 0.48$ & \textcolor{red}{\xmark}\textcolor{red}{\xmark} & $3.9 \, \pm \, 1.6$ \textcolor{ForestGreen}{\pmb{\checkmark}} \\[0,1cm]
        Boson star 3 & $0.99 \, \pm \, 0.22$ & $66 \, \pm \, 11$ & $0.16 \, \pm \, 0.13$ & \textcolor{red}{\xmark}\textcolor{red}{\xmark} & $1.4 \, \pm \, 0.9$ \textcolor{red}{\xmark} \\[0,1cm]
        Fluid sphere 2 & $1.01 \, \pm \, 0.13$ & $68 \, \pm \, 7$ & $0.01 \, \pm \, 0.32$ & \textcolor{red}{\xmark}\textcolor{red}{\xmark} & $0.81 \, \pm \, 0.45$ \textcolor{red}{\xmark}\\[0,1cm]
        Fluid sphere 3 & $0.91 \, \pm \, 0.23$ & $63 \, \pm \, 11$ & $0.20 \, \pm \, 0.47$ & \textcolor{red}{\xmark}\textcolor{red}{\xmark} & $1.7 \, \pm \, 1.2$ \textcolor{red}{\xmark}\\[0,1cm]
        Gravastar 1 & $1.00 \, \pm \, 0.13$ & $67 \, \pm \, 7$ & $0.18 \, \pm \, 0.16$ & \textcolor{red}{\xmark}\textcolor{red}{\xmark} & $1.7 \, \pm \, 0.5$ \textcolor{orange}{\boldmath$\sim$\unboldmath}\\[0,1cm]
        Gravastar 2 & $0.97 \, \pm \, 0.25$ & $65 \, \pm \, 12$ & $0.13 \, \pm \, 0.09$ & \textcolor{red}{\xmark}\textcolor{red}{\xmark} & $2.2 \, \pm \, 1.1$ \textcolor{orange}{\boldmath$\sim$\unboldmath}\\[0,1cm]
        Gravastar 3 & $1.01 \, \pm \, 0.21$ & $68 \, \pm \, 11$ & $0.02 \, \pm \, 0.20$ & \textcolor{red}{\xmark}\textcolor{red}{\xmark} & $1.0 \, \pm \, 0.7$ \textcolor{red}{\xmark}\\[0,1cm]
        \hline
    \end{tabular}
    \caption{Fit summary results for the studied metrics and Bayesian criteria: The first column lists the metric used for simulated data. The second column report the average best reduced chi-squared $<\chi^2_\mathrm{red}>$ with its standard deviation. The third column shows the average associated Bayesian Information Criterion with its standard deviation. The fourth column shows the average BIC-based Bayes factor and its standard deviation. The fifth column indicates the background metric detectability according to the criteria defined in~\ref{sec:detectability}. The final column reports the BIC-based Bayes factor between the best-fitting metric and Kerr, and assesses the exclusion of the latter. The uncertainties incorporated in the simulated data reflect GRAVITY's typical uncertainties.}
    \label{tab:detec_metric_current_uncertainties}
\end{table*}

\begin{figure*}[h]

    \centering
    \includegraphics[width=\linewidth]{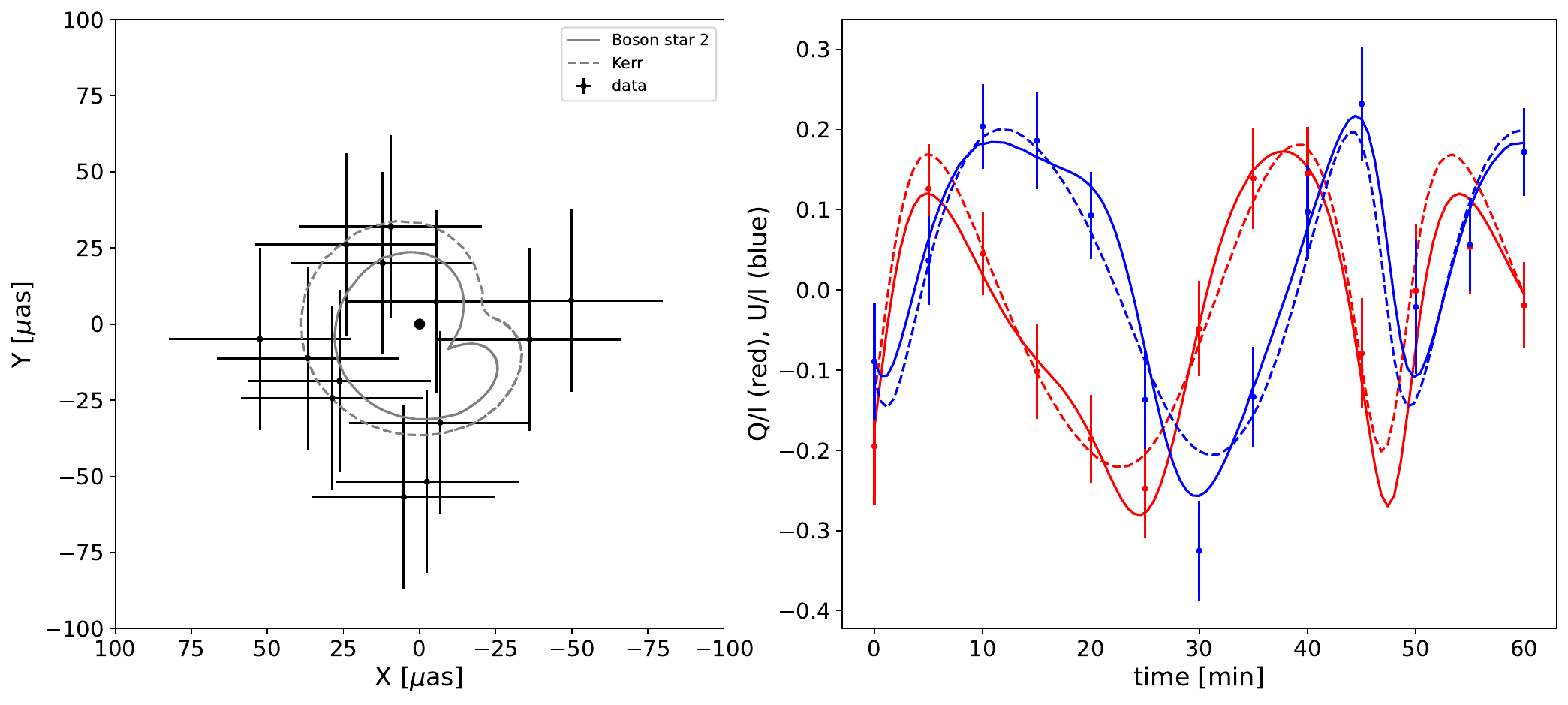}
    \captionof{figure}{Simulated data, generated in the Boson star 2 metric and represented by dots with error bars (reflecting GRAVITY-like uncertainties), are compared with two best-fitted models: one in the Boson star 2 metric (solid line) and one in the Kerr metric (dashed lines). \textbf{Left-panel} shows the astrometry, and \textbf{right-panel} shows the time evolution of $Q/I$ in red and $U/I$ in blue.}
    \label{fig:compareBS2_Kerr}

    \vspace{1em}
        
    \centering
    \begin{tabular}{l c c c c c c c}
      \hline
      \hline
      Metric & $\chi^2_\mathrm{red}$ & $r$ [$r_g$] & $i$ [$\degree$] & $\varphi_0$ [$\degree$] & $lp$ & $K_\mathrm{coef}$ & spin \\[0,1cm]
      \hline
      Boson star 2 & 0.66 & $8.05 \pm 0.28$ & $157.0 \pm 1.1$ & $10.0 \pm 2.4$ & $0.41 \pm 0.03$ & $1.03 \pm 0.06$ & 0 (fixed) \\[0,1cm]
      Kerr & 0.98 & $7.66 \pm 0.43$ & $160.4 \pm 1.5$ & $6.1 \pm 3.8$ & $0.31 \pm 0.03$ & $0.99 \pm 0.08$ & $0.29 \pm 0.30$ \\[0,1cm]
      \hline
    \end{tabular}
    \captionof{table}{Summary of the results of the fitting of one of the simulated data generated in the Boson star 2 metric (with GRAVITY uncertainties) for the best-fitting metric (the Boson star 2 model) and Kerr. These two models are shown with the simulated data in the figure above.}
    \label{tab:results_fits}
\end{figure*}

We first tested the detectability of the studied ECO metrics with current GRAVITY uncertainties. We generated 10 simulated data with $\sigma_\mathrm{Astro}=30\, \mu$as and the $\sigma_\mathrm{Flux}$ from Eq.~\ref{eq:sigma_flux} in all previously listed background metrics. 
For each of the generated datasets, we performed a fit in all background metrics, whose results are listed in Table~\ref{tab:detec_metric_current_uncertainties}. This table shows for each simulated data generated in a given background metric (each line), the average best reduced chi-squared $<\chi^2_\mathrm{red}>$ with its standard deviation (second column), the average associated $\mathrm{BIC}$ with its standard deviation (third column), the average and standard deviation of the logarithmic BIC-based Bayes factor $<\log_{10} K>$ (forth column). Columns five and six highlight the detectability of the metric used to generate the simulated data compared to all other metrics and compared only to Kerr resp., according to the criteria listed in~\ref{sec:detectability}. Column six also reports the average BIC-based Bayes factor between the best-fitting metric and Kerr, with its standard deviation.

The first main conclusion from Table~\ref{tab:detec_metric_current_uncertainties} is that none of the metric signatures can be clearly detected from the other with GRAVITY uncertainties ($<\log_{10} K>$ is always below 1). This is an expected result as, so far, all tests made are compatible with the Kerr solution. Moreover, in all cases, there is a risk of mismatch between the background metric and the best-fitting one. For most metrics, with the exception of Boson Star 2, across the 10 simulated datasets generated from a given background metric, none of the fitted metrics clearly outperforms the others, and no clear trend emerges. In other words, at this level of flux uncertainty, the mismatch does not occur between a particular pair of metrics but rather affects all candidate metrics. This arises because the impact of the metric (e.g., spin in the Kerr case, or plunge-through images for ECOs) on the astrometric and polarimetric observables is not strong enough relative to the uncertainties to break the degeneracies with the hot-spot parameters. Slightly different hot-spot parameters in two distinct metrics can thus produce observables that agree within the error bars.

However, as previously noted, this behavior is not seen for the Boson Star 2 model. Among the 10 simulated datasets, there is only one case where the Boson Star 2 is not the best-fitting metric; there, the data are better fitted by the Grava Star 1 metric, with $\chi^2_\mathrm{red}=1.22$ and $<\log_{10} K>=0.05$. In that dataset, the fit quality for Grava Star 1 and Boson Star 2 is therefore essentially the same. In all other cases, we clearly recover the Boson Star 2 metric. This behavior arises from the large angular size and relative brightness of the plunge-through images in this model (see top-left panel of Fig.~\ref{fig:BosonStar_Fluid_Sphere}), which strongly affect the observed polarization fraction and EVPA~\citep{Rosa2025}. This is illustrated in Fig.~\ref{fig:BS2_Contrib}, which shows the contributions of the different image orders and types: primary only (dotted), primary + secondary (equivalent to Schwarzschild; dashed), and all images including the plunge-through (solid). The contribution of the plunge-through images varies with time and orbital phase. This signal is strong enough that it cannot be mimicked by the spin of a Kerr black hole, as indicated by the BIC-based Bayes factor between the Boson star and Kerr fits of $<\log_{10} K|_\mathrm{Kerr}> = 3.9 \, \pm \, 1.6$. Thus, even with GRAVITY uncertainties, if Sgr A* is a Boson star (Boson Star 2 model), the data should still be better described by this model than by Kerr. This is illustrated in Fig.~\ref{fig:compareBS2_Kerr}, which shows the fits to one simulated dataset with a Boson Star 2 background, using both the Boson Star 2 and Kerr metrics. The fitted values are summarized in Table~\ref{tab:results_fits}.

\subsection{Spin vs. ECO signatures}\label{sec:test_spin}

\begin{figure}[h]
    \centering
    \includegraphics[width=\linewidth]{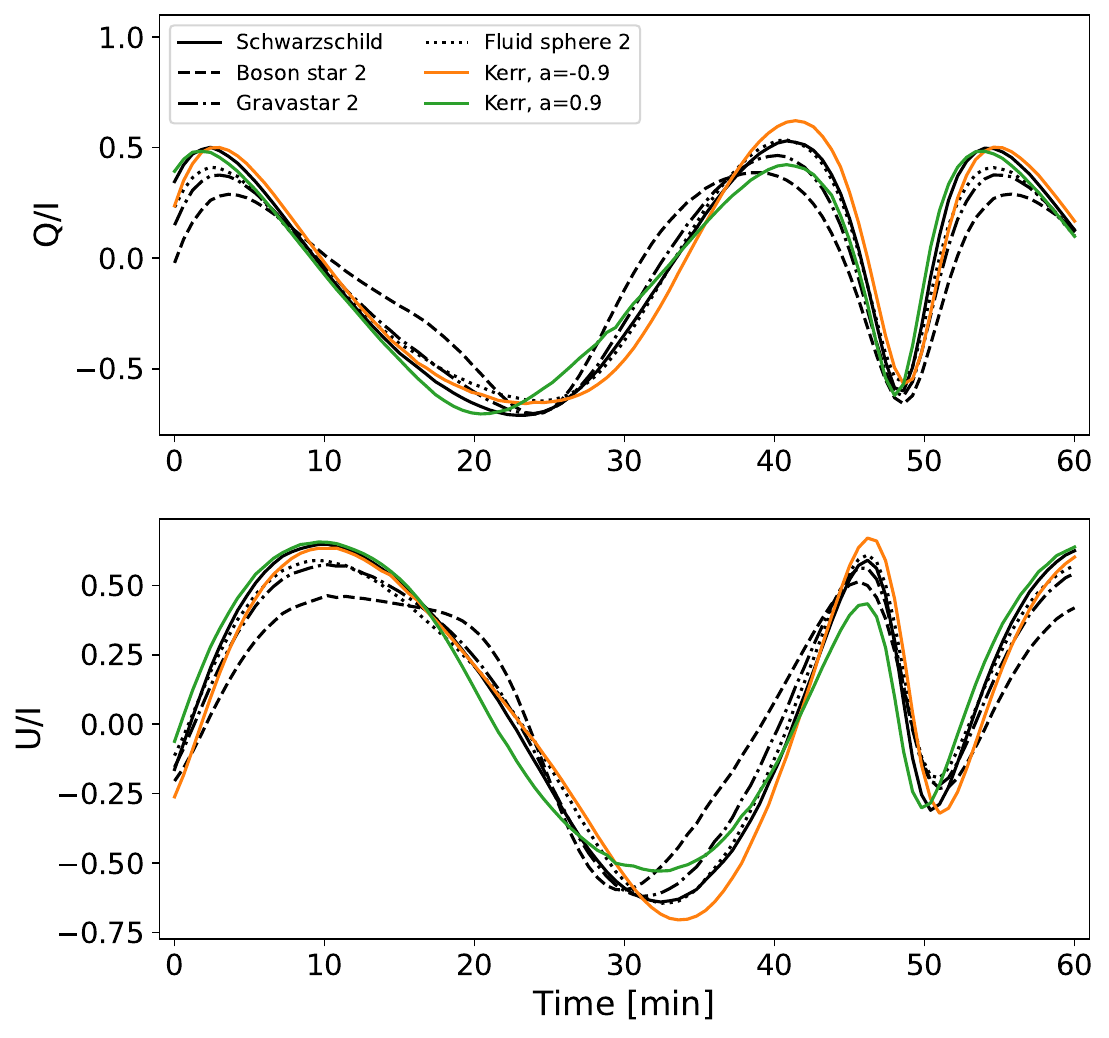}
    \caption{Normalized polarized fluxes of a hot-spot orbiting various compact metric models: Schwarzschild in black full lines, the Boson star 2 model in dashed lines, the Fluid sphere 2 model in dotted lines, the Gravastar 2 model in dashed-dotted lines, and two Kerr metrics with spin value of $a=-0.9$ in orange and $a=0.9$ in green. The parameters for the hot-spot are listed in Table~\ref{tab:hotspot_params}.}
    \label{fig:compare_spin}
\end{figure}

A central point of our analysis is the possibility that the polarization arising from plunge-through images in horizonless ECO metric models studied can mimic the effects of the spin of a Kerr black hole (and conversely). In Fig.~\ref{fig:compare_spin}, we present the normalized polarization fluxes $Q/I(t)$ and $U/I(t)$ over one orbit, using the hot-spot parameters from Table~\ref{tab:hotspot_params}, for various metrics, including ECOs and Kerr spacetimes with spins $a=-0.9$ (orange) and $a=0.9$ (green). The signal amplitudes relative to the Schwarzschild case (solid black curves) are broadly comparable between the spinning black holes and the ECO models (with the exception of the Boson star 2 model, as noted previously). However, the time dependence of the polarization differs across the various metrics, which in principle allows them to be distinguished observationally.

\subsubsection{Fitting an ECO with Kerr}
In this section, we examine how the fitting procedure is affected when applied to a simulated dataset generated in an ECO model with the Kerr metric.

Due to the contribution of plunge-through images, the astrometric signal is shrunk compared to the Schwarzschild case. Consequently, when fitting simulated data (astrometry + polarimetry) generated in an ECO model with the Kerr metric, the inferred orbital radius is generally expected to be smaller than the true input radius. This trend is observed in most case, however, the large astrometric uncertainties prevent us from placing tight constraints on this parameter.

\begin{figure}
    \centering
    \includegraphics[width=\linewidth]{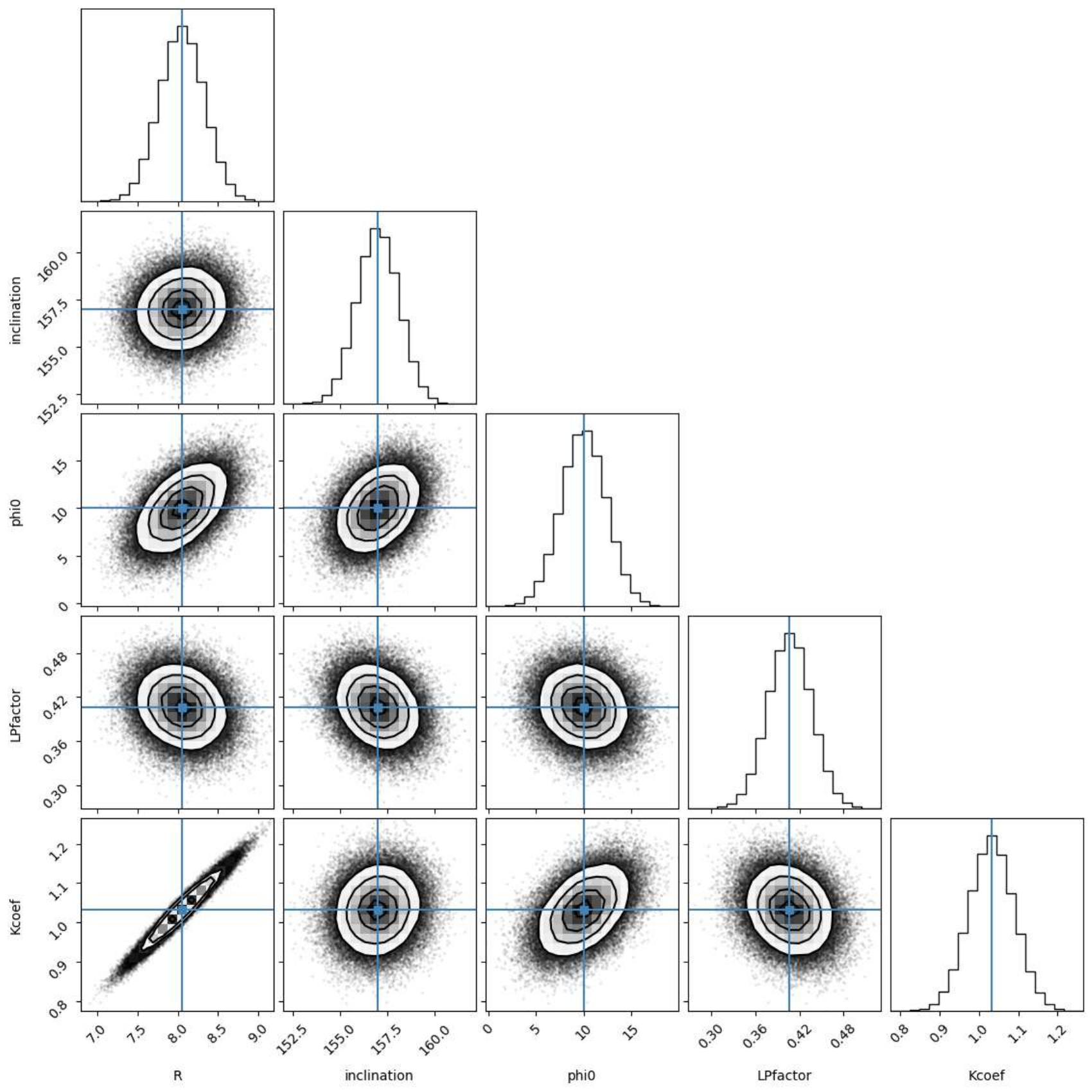}
    \caption{Corner plot illustrating the parameter fit for a single simulated dataset, generated and fitted in the Boson star 2 metric (corresponding to Table~\ref{tab:results_fits} and full line of Fig.~\ref{fig:compareBS2_Kerr}).}
    \label{fig:corner}
\end{figure}

In addition, the orbital velocity scaling factor $K_\mathrm{coef}$ is influenced by the value of the orbital radiusand thus separates the spatial from the temporal characteristics of the flares. Because astrometric data provide only weak spatial constraints, whereas polarimetric observations impose strong temporal constraints, changes in the orbital radius translate directly into changes in the inferred $K_\mathrm{coef}$. Consequently, the two parameters are strongly correlated, as illustrated by the corner plot in Fig.~\ref{fig:corner}.

We observed that when simulated data generated in an ECO model are fitted using a Kerr metric, the inferred linear polarization factor is significantly reduced relative to its true input value. Consequently, a low measured polarization fraction could falsely suggest the presence of plunge-through images. Nonetheless, many other poorly constrained astrophysical processes can also modify the polarization fraction. This introduces a complete degeneracy between an ECO-related origin and conventional astrophysical effects, and therefore we do not consider this behavior to be a reliable signature for identifying ECOs.

Finally, as discussed in Sect.~\ref{sec:params_summary}, when fitting any simulated data in the Kerr metric, the black hole spin is treated as a free parameter. We thus allow the algorithm to exploit this additional degree of freedom when fitting signals that include plunge-through images produced by ECOs. Nevertheless, in approximately $\sim 83 \%$ of the cases (58 out of 70) with GRAVITY-level uncertainties, the best-fit spin for data generated in an ECO metric is statistically consistent with zero. This behavior is illustrated by the fits reported in Table~\ref{tab:results_fits}. This indicates that varying the spin alone cannot fully reproduce the effects induced by plunge-through images.

Consequently, we examine if the datasets are best fitted by an ECO or Kerr and if we can exclude the latter. We employ the same statistical diagnostic as in Sect.~\ref{sec:methodology}, namely the BIC-based Bayes factor comparing the best-fitting metric to Kerr ($<\log_{10} K|_\mathrm{Kerr}>$). As previously mentioned, for the Boson star 2 metric, the corresponding model is statistically preferred over Kerr, whereas this does not hold for the other metrics when GRAVITY-like uncertainties are assumed. The Gravastar 1 and Gravastar 2 metrics yield a BIC-based Bayes factor with respect to Kerr ($\log_{10} K|_\mathrm{Kerr}$) exceeding one, but still below the threshold (once uncertainties are taken into account) required to draw a strong conclusion.

\subsubsection{Fitting Schwarzschild with an ECO vs with Kerr}
We now investigate the opposite configuration, namely fitting the simulated dataset in Schwarzschild using an ECO model and a Kerr model (see the first row of Table~\ref{tab:detec_metric_current_uncertainties}).

The first thing to mention is that for five simulated datasets out of ten, with GRAVITY uncertainties and based on $\chi^2_\mathrm{red}$, we recovered the correct background metric, but for the other five, there has been a mismatch. When we retrieved the black hole metric, the BIC value is higher compared to the one of the second best-fitting metric because of the extra degree of freedom allowed in Kerr, i.e. the spin. This results in a negative $\Delta \mathrm{BIC}$ and consequently a negative $\log_{10} K$, explaining the negative value of $<\log_{10} K>$. This means that even if the best-fitting metric is Kerr according to the $\chi^2_\mathrm{red}$, this is only due to the spin being free and that this fitting is not significantly better - and actually equivalent - to the fits in ECOs models.

Another interesting result is when we look at the BIC-based Bayes factor relative to Kerr ($<\log_{10} K|_\mathrm{Kerr}>$). Indeed, this value is compatible with zero but has significant error bars. There is even an extreme case for which the simulated data are best-fitted in the Fluid Sphere 2 model and has a $\log_{10} K|_\mathrm{Kerr}=2.01$ which, in case of a real flare, could be accounted for a detection of the later metric. This highlights the fact that with these uncertainties, there is a risk of false detection. The results of this fit and the comparison with the best-fitting in Kerr are shown in Appendix~\ref{ap:fit_Kerr_by_FS2}.

\subsection{Improve the detectability of the ECOs} %Future uncertainties with GRAVITY+/VLTI upgrades
\label{sec:future_uncertainties}
In 2026, the upgrade of the GRAVITY instrument called \href{https://www.eso.org/public/teles-instr/paranal-observatory/vlt/vlt-instr/gravity+/}{GRAVITY+} should enter in service. This upgrade of the instrument itself also comes with an improvement of the VLTI infrastructure, including a better Adaptive Optic (AO) system with laser guide stars on all Unit Telescopes (UT; 8 meters telescopes) and better fringe tracking capabilities. The anticipated outcome of these enhancements is, optimally, an improvement of a factor of~$\sim7$ in the signal-to-noise ratio with respect to flux uncertainties~\citep{Bourdarot2024}.

We then investigated how one can improve the detectability of the studied ECOs models in the context of the upcoming GRAVITY+ upgrade. The first obvious possibility is to reduce the uncertainties. While smaller flux uncertainties is achievable, reducing the astrometric uncertainties are challenging as GRAVITY is at the state of the art for astrometric precision. We thus focused on improvement on flux sensitivity and investigated two scenarios: 1) smaller flux uncertainties with the same temporal resolution (Sect.~\ref{sec:smaller_flux_uncertainties}), and 2) similar flux uncertainties with higher temporal resolution (Sect.\ref{sec:better_time_sampling}).

\subsubsection{Smaller flux uncertainties} \label{sec:smaller_flux_uncertainties}
We therefore investigate the best improvement on flux uncertainties expected by the GRAVITY+ upgrade. We performed the same analysis as in Sect.~\ref{sec:current_uncertainties} with a seven times smaller uncertainty of the flux $\sigma_\mathrm{Flux}$. 

Given the high level of photometric precision achieved, the impact of the plunge-through images, as induced by the various ECO models, exceeds the corresponding uncertainty. As a result, the data are fitted significantly better when using the same metric adopted to generate the simulated data, compared with alternative metrics (see Table~\ref{tab:detec_metric_uncertainties_GRAVITY+}). This effect is especially pronounced for the Schwarzschild, Boson star 2, and Fluid sphere 2 metrics, all of which satisfy the detectability conditions (zero mismatch and $<\log_{10} K>$ greater than 2). The Boson star 3 model also shows enhanced detectability; however, its mean BIC-based Bayes factor minus one standard deviation remains below the threshold value of two needed to assert detectability. 

\begin{table*}[ht]
   \centering
    \begin{tabular}{l|c c c c c}
        \hline
        \hline
        Data's metric & $< \chi^2_\mathrm{red}>$ & $<BIC>$ & $<\log_{10} K>$ & detectability & $<\log_{10} K|_\mathrm{Kerr}>$ \\[0,1cm]
        \hline
        Schwarzschild & $1.09 \, \pm \, 0.21$ & $74 \, \pm \, 10$ & $11.12 \, \pm \, 3.78$ & \textcolor{ForestGreen}{\pmb{\checkmark}} & \\[0,1cm]
        Boson star 2 & $0.95 \, \pm \, 0.14$ & $64\, \pm \, 7$ & $22 \, \pm \, 5.9$ & \textcolor{ForestGreen}{\pmb{\checkmark}} & $125 \, \pm \, 8$ \textcolor{ForestGreen}{\pmb{\checkmark}} \\[0,1cm]
        Boson star 3 & $0.88 \, \pm \, 0.25$ & $61 \, \pm \, 12$ & $3.0 \, \pm \, 1.4$ & \textcolor{orange}{\boldmath$\sim$\unboldmath} & $22 \, \pm \, 2.7$ \textcolor{ForestGreen}{\pmb{\checkmark}} \\[0,1cm]
        Fluid sphere 2 & $ 1.00\, \pm \, 0.22$ & $67 \, \pm \, 11$ & $6.5 \, \pm \, 1.8$ & \textcolor{ForestGreen}{\pmb{\checkmark}} & $15 \, \pm \, 6$ \textcolor{ForestGreen}{\pmb{\checkmark}}\\[0,1cm]
        Fluid sphere 3 & $1.09 \, \pm \, 0.14$ & $71 \, \pm \, 7$ & $2.1 \, \pm \, 1.3$ & \textcolor{red}{\xmark}\textcolor{red}{\xmark} & $38 \, \pm \, 6$ \textcolor{ForestGreen}{\pmb{\checkmark}}\\[0,1cm]
        Gravastar 1 & $1.08 \, \pm \, 0.13$ & $70 \, \pm \, 6$ & $1.8 \, \pm \, 0.75$ & \textcolor{red}{\xmark}\textcolor{red}{\xmark} & $22 \, \pm \, 5$ \textcolor{ForestGreen}{\pmb{\checkmark}}\\[0,1cm]
        Gravastar 2 & $0.98 \, \pm \, 0.17$ & $66 \, \pm \, 8$ & $0.68 \, \pm \, 0.41$ & \textcolor{red}{\xmark}\textcolor{red}{\xmark} & $31 \, \pm \, 4$ \textcolor{ForestGreen}{\pmb{\checkmark}}\\[0,1cm]
        Gravastar 3 & $1.21 \, \pm \, 0.15$ & $77 \, \pm \, 7$ & $3.3 \, \pm \, 2.0$ & \textcolor{red}{\xmark}\textcolor{red}{\xmark} & $11 \, \pm \, 6$ \textcolor{ForestGreen}{\pmb{\checkmark}} \\[0,1cm]
        \hline
    \end{tabular}
    \caption{Same as Table~\ref{tab:detec_metric_current_uncertainties} with error bars on flux $7$ times better compared to GRAVITY typical uncertainty, which corresponds to the improvement of sensitivity of GRAVITY+.}
    \label{tab:detec_metric_uncertainties_GRAVITY+}
\end{table*}

\begin{figure*}[h]
    \centering
    \includegraphics[width=\linewidth]{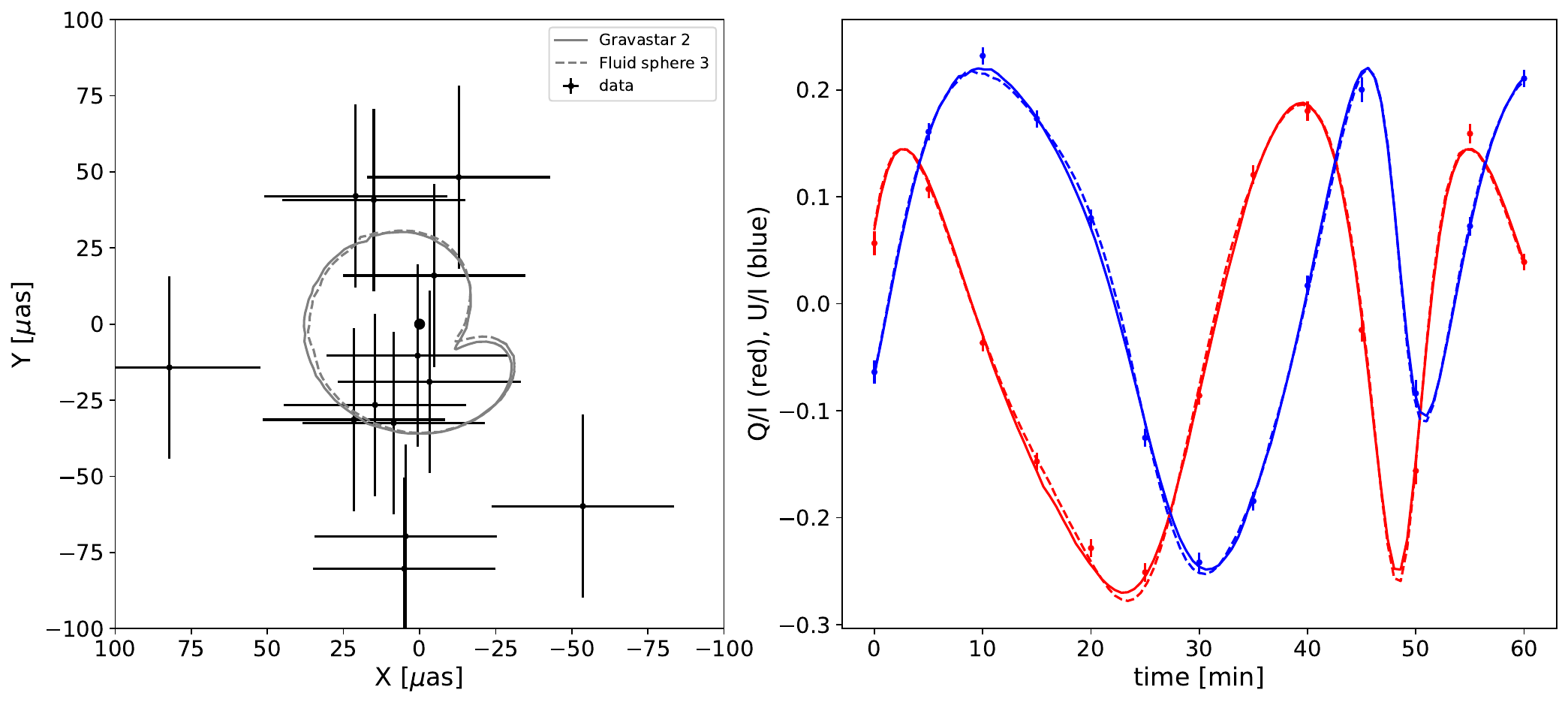}
    \caption{Same as Fig.~\ref{fig:compareBS2_Kerr} with simulated data in the Fluid sphere 3 background and best GRAVITY+ uncertainties. The best-fitting models shown are in the Gravastar 2 metric (full lines) and in the Fluid sphere 3 metric (dashed lines).}
    \label{fig:mismatch_FS3_GS2_GRAVITY+}

    \vspace{1em}

    \begin{tabular}{l c c c c c c c}
      \hline
      \hline
      Metric & $\chi^2_\mathrm{red}$ & $r$ [$r_g$] & $i$ [$\degree$] & $\varphi_0$ [$\degree$] & $lp$ & $K_\mathrm{coef}$ & spin \\[0,1cm]
      \hline
      Gravastar 2 & 0.98 & $8.17 \pm 0.08$ & $157.9 \pm 0.2$ & $15.7 \pm 0.7$ & $0.39 \pm 0.01$ & $0.99 \pm 0.02$ & 0 (fixed) \\[0,1cm]
      Fluid sphere 3 & 1.09 & $8.35 \pm 0.07$ & $157.4 \pm 0.2$ & $16.4 \pm 0.6$ & $0.40 \pm 0.01$ & $1.02 \pm 0.02$ & 0 (fixed) \\[0,1cm]
      \hline
    \end{tabular}
    \captionof{table}{Summary of the results of the fitting of one of the simulated data generated in the Fluid sphere 3 with GRAVITY+ uncertainties for the best-fitting metric (the Gravastar 2 model) and in the Fluid sphere 3 model.}
    \label{tab:results_fits_FS3_GS2}
\end{figure*}

The remaining metrics—namely Fluid sphere 3 and the three Gravastar configurations—continue to exhibit mismatches even under these small uncertainties. In the case of Fluid sphere 3, only two realizations show a mismatch, specifically with the Boson star 3 and Gravastar 2 models. However, Gravastar 2 is most frequently identified as the second-best fit. We further observe that when the simulated datasets are generated in the Gravastar 2 background, the most common mismatches occur with the Fluid sphere 3 model. This indicates that, from an observational standpoint, the signals produced by Fluid sphere 3 and Gravastar 2 are very similar and remain close enough to each other to induce confusion between the two, even at such low uncertainty levels. In Fig.~\ref{fig:mismatch_FS3_GS2_GRAVITY+}, we present the best-fitting models for these two metrics (Fluid sphere 3 shown as solid lines and Gravastar 2 as dashed lines) for a single simulated dataset generated in the Fluid sphere 3 metric that leads to a mismatch. This figure clearly shows that two metrics can have the same observational properties, with tiny differences leading to a mismatch.
For the Gravastar 1 configuration, there are once more only two mismatches with the Gravastar 2 and Boson star 3 models. As in the case of the Fluid sphere 3 model, Gravastar 2 again emerges as the metric that most frequently provides the second-best fit. This is expected given that the time-integrated images of the Gravastar 2 and 3 models shown in Fig.~\ref{fig:GravaStar} are similar. To enhance the accuracy and detectability of these models, incorporating the radius of the Gravastar as a free, continuous parameter in the fitting process (an approach that exceeds the boundaries of this paper) rather than selecting discrete values could prove beneficial.

\begin{figure*}
    \centering
    \includegraphics[width=\linewidth]{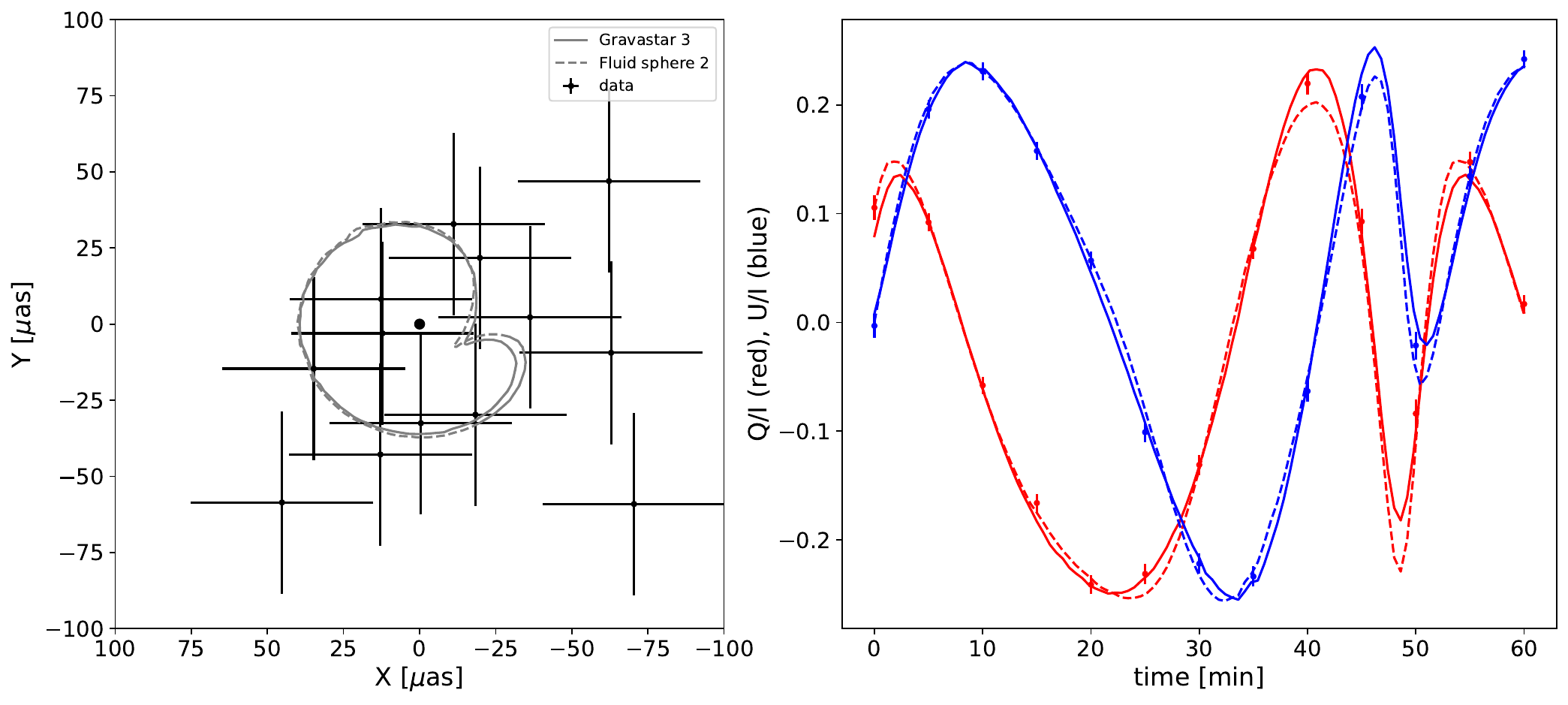}
    \caption{Same as Fig.~\ref{fig:compareBS2_Kerr} with simulated data in the Gravastar 3 background and best GRAVITY+ uncertainties. The best-fitting models shown are in the Gravastar 3 metric (full lines) and in the Fluid sphere 2 metric (dashed lines).}
    \label{fig:compareGS3_FS2_GRAVITY+}

    \vspace{1em}

    \begin{tabular}{l c c c c c c c}
      \hline
      \hline
      Metric & $\chi^2_\mathrm{red}$ & $r$ [$r_g$] & $i$ [$\degree$] & $\varphi_0$ [$\degree$] & $lp$ & $K_\mathrm{coef}$ & spin \\[0,1cm]
      \hline
      Gravastar 3 & 1.23 & $8.14 \pm 0.04$ & $157.1 \pm 0.3$ & $15.0 \pm 0.7$ & $0.42 \pm 0.01$ & $0.99 \pm 0.01$ & 0 (fixed) \\[0,1cm]
      Fluid sphere 2 & 1.32 & $8.35 \pm 0.09$ & $155.7 \pm 0.3$ & $15.4 \pm 0.6$ & $0.40 \pm 0.01$ & $1.02 \pm 0.02$ & 0 (fixed) \\[0,1cm]
      \hline
    \end{tabular}
    \captionof{table}{Summary of the results of the fitting of one of the simulated data generated in the Gravastar 3 with GRAVITY+ uncertainties for the best-fitting metric (the Gravastar 3 model) and in the Fluid sphere 2 model.}
    \label{tab:results_fits_GS3_FS2}
\end{figure*}

The case of the Gravastar 3 is different from the others. Indeed, even with the low uncertainties, there is a significant amount of mismatch (40\%) with the Fluid sphere 2 models. When we recover the correct background metric, the second best-fitting metric is always the Fluid sphere 2 model, suggesting similar observational properties. However, the opposite, i.e., when the simulated datasets are in the Fluid sphere 2, this trend of degeneracies is not observed.
We show in Fig.~\ref{fig:compareGS3_FS2_GRAVITY+} one simulated dataset generated in the Gravastar 3 model with the best-fit made in the latter metric in full lines and in the Fluid sphere 2 model in dashed lines. The associated best-fitted parameters are listed in Table~\ref{tab:results_fits_GS3_FS2}. The two signals are very similar and indistinguishable at early times but start to differ at later times between $t=35$ min and $t=55$ min, corresponding to $\varphi \sim 225 \degree$ and $\varphi \sim 345 \degree$ respectively. In this region, the primary image is dimmed because of the small pitch angle between the photon wave vector and the ambient vertical magnetic field (lower synchrotron emission). As a consequence, the signals are dominated by the secondary and plunge-through images, explaining the higher difference between the two models.

The most important result is that with the best GRAVITY+ improvement on flux uncertainties, for all simulated datasets in an ECO metric, the mean BIC-based Bayes factor with Kerr ($<\log_{10} K|_\mathrm{Kerr}>$) minus one standard deviation is higher than two. This means that if the background metric is one of the studied ECOs, even if we cannot fully constrain the nature of the ECO (it depends on the model), we can exclude the Kerr metric. We thus will be able to test the Kerr hypothesis using Sgr~A*'s flare polarization and astrometry.  

We note that these high polarimetric measurements are also achievable by the radio interferometer Atacama Large Millimeter/submillimeter Array (ALMA). However, the environmental effects by the surrounding plasma at radio wavelengths are stronger than in IR and there is no possible astrometry because of angular resolution.

\subsubsection{Better time resolution} \label{sec:better_time_sampling}

An alternative utilization of the GRAVITY+ and VLTI enhancements involves reducing the integration time from the standard 5-minute intervals to a more frequent 1-minute interval, while maintaining flux uncertainties comparable to those observed with the GRAVITY measurements. For observation of flares at the galactic center, such a scenario is more likely than the improvement of a factor seven on the flux uncertainties.

The impact of plunge-through images on polarization is not constant over time as shown by Figs~\ref{fig:BS2_Contrib}~and~\ref{fig:compare_spin}, suggesting that enhanced temporal resolution, achieved without compromising flux uncertainty, might be enough to differentiate between different metrics effectively. To evaluate this hypothesis, we have created simulated datasets, capturing one data point every minute, repeating the prior procedures to examine the results. However, the fit results for each dataset are broadly consistent with those obtained using the 5-minute cadence, with most values of $<\log_{10} K>$ (again with the exception of BS2) remaining below the threshold of unity. This occurs because, although the metric’s signature is time dependent, the substantial flux uncertainties prevent us from clearly disentangling the metric signal from the astrophysical parameters. Consequently, this method does not yield a definitive conclusion, and the most effective strategy is instead to reduce the flux uncertainties.

\section{Limitations and discussion}\label{sec:discussion}
\subsection{Limitations}
In this study, we adopt a simplified hot-spot model characterized by uniform spatial parameters positioned along a circular orbit within the equatorial plane at the Schwarzschild-Keplerian velocity. This model is likely too rudimentary to adequately capture the flares of Sgr A* in NIR observations. The nature of motion, whether circular, helical, or conical, significantly influences both the observed astrometry~\citet{Antonopoulou2024} and the light curves~\citep{Aimar2023}. The electron energy distribution (EED) is presumed to follow a Kappa distribution, with a constant power-law index at high energies, number density, and temperature, while variability is highly expected for flares. Indeed, acceleration mechanisms like magnetic reconnection, for example, and synchrotron cooling play a crucial role in high-energy electrons and therefore flare modeling.

With a more sophisticated model, which includes more astrophysics (with its uncertainties), it is probable that the detectability of the metrics will decrease as the space-time characteristics could be diluted and degenerated with the astrophysical parameters. This highlights the importance of accurate flare modeling.

Moreover, until now, we have not accounted for the quiescent state of Sgr~A*. Although its impact is low when the flux from the flare is high, when the flux from the flare and the quiescent are comparable, i.e. at a later time, the quiescent might not be neglected.

\subsection{Discussion}
The encouraging results outlined in the previous discussion hold significant implications for the future observations planned with the advanced GRAVITY+ instrument and the upgraded Very Large Telescope Interferometer (VLTI). These advancements aim to detect the elusive ECO signatures in the flares emitted by Sagittarius A* (Sgr~A*). Additionally, analogous polarization studies of Sgr~A* flares have been conducted at radio frequencies, as detailed by~\citet{Wielgus2022}. At these wavelengths, considerations such as Faraday rotation and conversion effects cannot be disregarded, as the phenomena of accretion and ejection remain intensely luminous even during flare events. This scenario adds further complexity to the astrophysical processes observed in comparison to those in the NIR spectrum. Despite these challenges, the exceptional temporal resolution provided by the Atacama Large Millimeter/submillimeter Array (ALMA) coupled with its remarkable ability to ascertain very low levels of polarized flux uncertainties, renders this instrument an exceedingly promising candidate for the rigorous search for ECO signatures within the radio flares of Sgr~A*.

The latest polarized images presented by~\citet{EHT_2024vii} provide a persuasive and distinct opportunity to explore the signatures of ECOs and to examine alternative gravitational theories~\citep{Yan2024, Ahmed2024, Aliyan2024, Walia2024, Vertogradov2025, Perrucci2025, Gan2024, Vishvakarma2025, Li2025}.

Ultimately, with the improvements brought by the GRAVITY+ upgrade and the enhancements of the VLTI, alongside the anticipated new generation of NIR instruments integrated with the Extremely Large Telescope (ELT), it is anticipated that an increased number of stars will be discerned~\citep{Bourdarot2024}, with some being in close proximity to Sgr~A*. This advance in observational capability will enable the conduction of further rigorous assessments of general relativity, the examination of ECO imprints on the trajectories of stellar bodies, as well as the evaluation of alternative gravitational theories. Such tests remain independent of the constraints posed by Sgr~A* polarized flares, offering a complementary approach.

\section{Summary and Conclusions}\label{sec:conclusion}
In this paper, we conducted an in-depth investigation into the detectability of space-time signatures through polarization measurements of individual flares from Sgr~A*. For our analysis, we employed an analytical model of a hot-spot orbiting the compact object in its equatorial plane, specifically at the Schwarzschild-Keplerian velocity. Our examination encompassed a comprehensive set of eight different background metrics, derived from four distinct families of compact objects: Schwarzschild black holes, Boson stars, Fluid spheres, and Gravastars. These models included parameters for exotic compact objects crafted to mimic the behavior of regular black holes. 

The principal distinction between the ECOs analyzed in this study and regular black holes lies in the existence of plunge-through images, which possess polarization properties comparable to the secondary image associated with the black hole space-time. These plunge-through images also contribute to an increased total intensity-which is, alone, completely degenerate with astrophysical parameters such as the number density or temperature of the hot-spot-thereby influencing both the observed EVPA and the polarization fraction. To break degeneracies and limit the number of the hot-spot parameters, it becomes imperative to examine the normalized polarized quantities $Q/I$ and $U/I$, rather than the original parameters $I$, $Q$ and $U$.

To evaluate the detectability of a specified metric, we generated ten simulated datasets in this metric and executed 8 fits (one for each metric under investigation) for each of them. The metric with the lowest $\chi^2_\mathrm{red}$ value is considered to be the best-fitting metric. To state whether a metric is detectable or not, we first check that the best-fitting metric corresponds to the one used to generate the simulated data (no mismatch). We then compare the mean BIC-based Bayes factor $<\log_{10} K>$ between the best-fitting metric and the other metrics. If the latter minus one standard deviation is greater than 2, we consider that the best-fits are sufficiently better than the others to state that the metric is detectable; otherwise, it is not. We also compare the best-fitting metric to the best-fit in the Kerr metric using the same threshold. This enables us to determine whether the data favor ECO models over Kerr metrics, or, equivalently, whether the Kerr metric is ruled out by the data or not.

The imprints on polarization measurements from the plunge-through images are, at best, of the same order as the current measurement uncertainties associated with the GRAVITY instrument. As a result, no metric is currently detectable (refer to Table~\ref{tab:detec_metric_current_uncertainties}). Nevertheless, for the Boson star 2 model, the best-fitting metric (most of the time the Boson star 2 model itself) is statistically significantly better compared to Kerr to exclude the latter. 
We also predicted the detectability of ECOs with the upcoming upgrade of the GRAVITY instrument: GRAVITY+. With a $\sim 7$ times smaller flux uncertainties than we can expect with this upgrade, some ECO metrics- the Boson star 2 and the Fluid sphere 2 models-becomes detectable, and the Boson star 3 model becomes potentially detectable with a limited BIC-based Bayes factor of $3.0 \pm 1.4$.
We examined whether an ECO model can mimic the spin signature of a Kerr black hole, and conversely whether a Kerr black hole can mimic an ECO. Although the signal amplitudes are broadly similar, their temporal characteristics differ, allowing the two scenarios to be discriminated provided the flux uncertainties are sufficiently good. In fact, with the best expected GRAVITY+ performance estimates, and for all ECO metrics considered, the average BIC-based Bayes factor with respect to Kerr ($<\log_{10} K|_\mathrm{Kerr}>$) exceeds the adopted threshold. Consequently, at this level of precision, if Sgr~A* were described by one of the ECO models analyzed, the application of this method would statistically prefer the ECO description over the Kerr solution.

However, the model used in this study is very simplistic, with very little astrophysics (only the vertical magnetic field configuration from~\citet{gravity2023} and synchrotron emission). Most of the photometric properties are governed by relativistic effects like beaming, Doppler boosting, and light bending. In reality, flares are much more complex, with many astrophysical uncertainties that mitigate detectability and could lead to an incorrect best-fit metric.

Nevertheless, this advancement opens up a realm of exciting opportunities and holds substantial promise for the future of astronomical observations, paving the way for innovative discoveries and pushing the boundaries of our understanding of the universe.

\begin{acknowledgements}
N.A. and P.G. thanks the Fundação para a Ciência e Tecnologia (FCT), Portugal, for the financial support to the Center for Astrophysics and Gravitation (CENTRA/IST/ULisboa) through grant No. UID/99/2025. The authors thankfully acknowledge the computer resources, technical expertise and assistance provided by CENTRA/IST. Computations were performed at the cluster “Baltasar-Sete-Sóis” and supported by the H2020 ERC Consolidator Grant "Matter and strong field gravity: New frontiers in Einstein's theory" grant agreement no. MaGRaTh-646597. J. L. R. is supported by the Project PID2022-138607NBI00, funded by MICIU/AEI/10.13039/501100011033 (``ERDF A way of making Europe" and ``PGC Generaci\'on de Conocimiento") The authors thank Saurabh, Thibaut Paumard and Frédéric Vincent for fruitful discussions. 
\end{acknowledgements}

\bibliographystyle{aa} % style aa.bst
\bibliography{biblio}

\appendix
\section{Grid search to find a good first-guess for Least-square} \label{ap:grid}
As said in the main text, the least-squares algorithm is sensitive to the first-guess. If one set the first-guess in a local minimum, the algorithm will be trapped in the latter. To avoid this situation, we computed a grid of models varying some of the hot-spot parameters, as the inclination, orbital radius, initial azimuthal angle and linear polarization fraction, in all background metrics. The grid characteristics are reported in~\ref{tab:grid}. We evaluate the $\chi^2_\mathrm{red}$ for all grid points, the one with the lowest $\chi^2_\mathrm{red}$ is selected as a first-guess for the least-squares algorithm in the associated metric. 

\begin{table}[h]
    \centering
    \begin{tabular}{l|cc}
        \hline
        \hline
        Parameter & Range & \# of grid points\\[0,1cm]
        \hline
        Inclination [$^\circ$] & $[135,170]$ & $8$ \\[0,1cm]
        Orbital radius [$r_g$] & $[8,12]$ & $9$\\[0,1cm]
        Initial Azimuthal angle [$^\circ$] & $[0,345]$ & $24$\\[0,1cm]
        Linear polarization factor & $[0,1]$ & $11$\\[0,1cm]
        \hline
    \end{tabular}
    \caption{Summary of the modeled grid used to find a good first-guess for the least-squares algorithm.}
    \label{tab:grid}
\end{table}

\section{Extreme mismatch between the Schwarzschild solution and an ECO-based model} \label{ap:fit_Kerr_by_FS2}
\begin{figure*}[h]
    \centering
    \includegraphics[width=\linewidth]{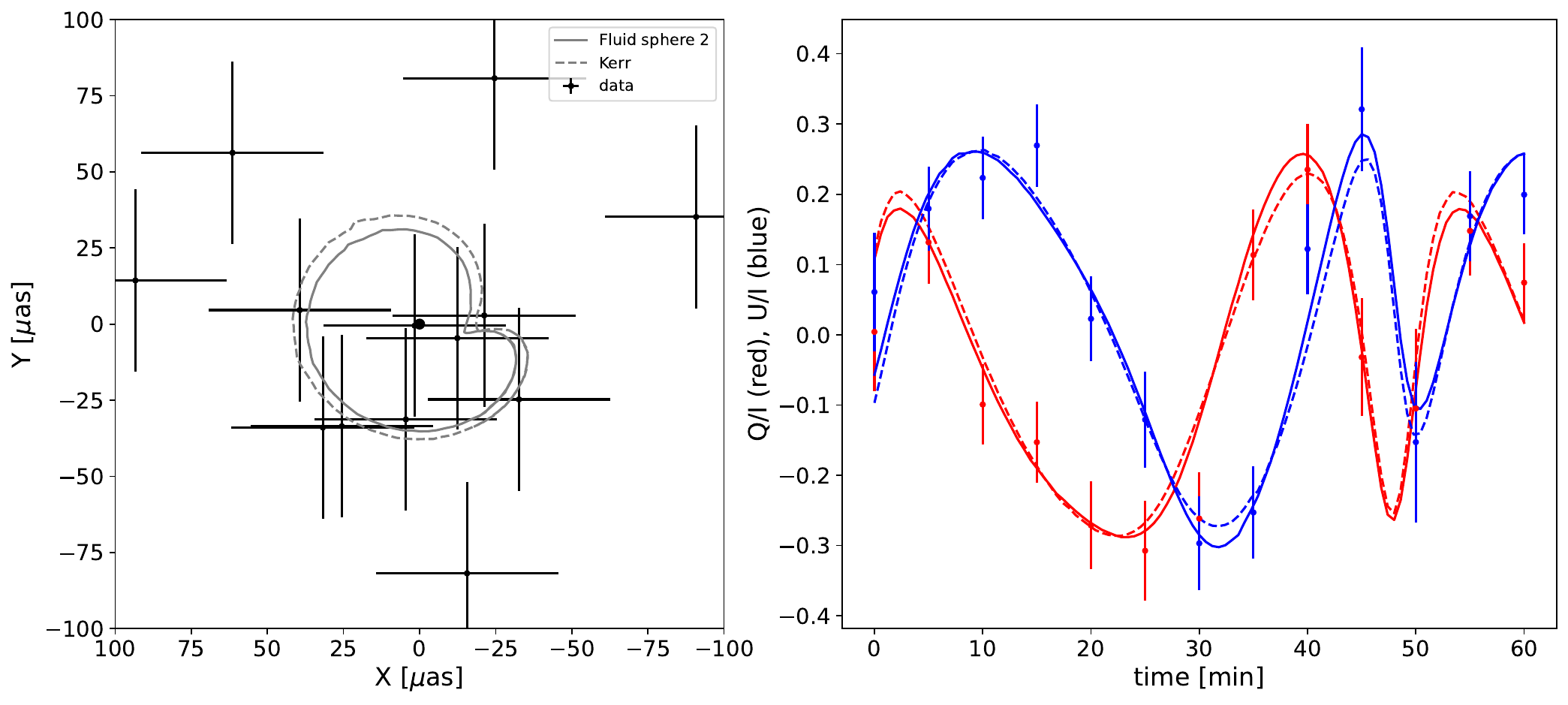}
    \captionof{figure}{Same as Fig.~\ref{fig:compareBS2_Kerr} with simulated data generated in Schwarzschild. The best-fitting metric is the Fluid sphere 2 model with parameters listed in Table~\ref{tab:results_fits_Sch_FS2} (full lines), and the best-fit model in Kerr is shown in dashed lines.}
    \label{fig:compareSch_FS2}

    \vspace{1em}

    \begin{tabular}{l c c c c c c c}
      \hline
      \hline
      Metric & $\chi^2_\mathrm{red}$ & $r$ [$r_g$] & $i$ [$\degree$] & $\varphi_0$ [$\degree$] & $lp$ & $K_\mathrm{coef}$ & spin \\[0,1cm]
      \hline
      Fluid sphere 2 & 1.13 & $7.85 \pm 0.36$ & $158.3 \pm 1.3$ & $15.7 \pm 3.6$ & $0.46 \pm 0.04$ & $0.94 \pm 0.07$ & 0 (fixed) \\[0,1cm]
      Kerr & 1.27 & $8.01 \pm 0.43$ & $157.9 \pm 1.5$ & $14.6 \pm 4.6$ & $0.41 \pm 0.04$ & $0.97 \pm 0.08$ & $-0.08 \pm 0.26$ \\[0,1cm]
      \hline
    \end{tabular}
    \captionof{table}{Summary of the results of the fitting of one of the simulated data generated in Schwarzschild (with GRAVITY uncertainties) for the best-fitting metric (the Fluid sphere 2 model) and in Kerr.}
    \label{tab:results_fits_Sch_FS2}
\end{figure*}

The result of the fitting of one of the simulated datasets generated in Schwarzschild with GRAVITY uncertainties has shown interesting results. Indeed, the best-fitting metric is the Fluid sphere 2 with a $\chi^2_\mathrm{red} = 1.13$ and is significantly better than the best-fit in the Kerr metric, which has a $\chi^2_\mathrm{red}= 1.27$. The BIC-based Bayes factor between the two is $\log_{10} K|_\mathrm{Kerr} = 2.01$ which, in real flare observation, could be interpreted as a detection of the Fluid sphere 2 model. This highlights the fact that with these uncertainties, there is a risk of false detection. Fig.~\ref{fig:compareSch_FS2} shows the best-fit in the Fluid sphere 2 model (full lines) and in Kerr (dashed lines) with the parameters listed in Table~\ref{tab:results_fits_Sch_FS2}.

\section{"Blind" test results}

\begin{table*}[ht]
    \centering
    \begin{tabular}{lc|cc|cc}
        \hline
        \hline
        Parameter & Symbol & Blind Test 1 & Best-fit & Blind Test 2 & Best-fit\\[0,1cm]
        \hline
        Orbital radius [$r_g$] & $r$ & $10$ & $10.07 \pm 0.07$ & $11$ & $10.68 \pm 0.05$\\[0,1cm]
        Azimuthal angle at $t^{BL}=0$ [°] & $\varphi_0$ & $180$ & $180.52 \pm 1.1$ & $5$ & $0.51 \pm 0.39$\\[0,1cm]
        Inclination [°] & $i$ & $160$ & $159.51 \pm 0.35$ & $148.1$ & $146.93 \pm 0.26$\\[0,1cm]
        Linear polarization factor & $lp$ & $0.7$ & $0.72 \pm0.02$ & $0.87$ & $0.85 \pm 0.01$\\[0,1cm]
        metric & & Gravastar 2 & Gravastar 2 & Fluid sphere 3 & Gravastar 1\\[0,1cm] 
        \hline
    \end{tabular}
    \caption{Summary of the parameters of the hot-spot model used to generate the simulated data for the blind tests and the fitted values.}
    \label{tab:hotspot_params_blindtest}
\end{table*}

To ensure the accuracy of our findings and verify that they do not rely on a particular selection of values, we have conducted two "blind" tests. In these tests, the parameters, including the background metric, remain unknown to the individual conducting the analysis. This approach effectively reduces any potential biases or preconceived notions that might influence the fitting process and its outcomes.

These tests involved the use of simulated data sets and, notably, these datasets were generated by individuals other than the one performing the data fitting analysis. Details concerning the parameters used to generate these simulated datasets, including the specific metric used, together with the resulting best-fit parameters, are reported in Table~\ref{tab:hotspot_params_blindtest}. We used the GRAVITY uncertainties for these tests. For the first blind dataset, which has been generated in the Gravastar~2 metric, we found the correct metric (and correct parameter estimation) as earlier, but the BIC-based Bayes factor is too low ($min(\log_{10} K) = 0.17$) to claim a detection (as before). For the second data set, which has been generated in the Fluid sphere 3 metric, we did not find the correct metric (it is the second-best metric) as the best-fitted metric is the Gravastar~1. We note that the estimated values of the parameters are a bit offset from the "real" values (especially $\varphi_0$). The BIC-based Bayes factor is again too low to distinguish the first and second-best fit metric ($\log_{10} K = 0.47$). However, in both cases, the Schwarzschild metric can be excluded. We thus obtained very similar results to those in Sect.~\ref{sec:current_uncertainties} validating our results.

\end{document}